\documentclass[12pt]{article}
\pdfoutput=1
\usepackage[utf8]{inputenc}
\usepackage[DIV=13]{typearea}
\usepackage{ragged2e}
\usepackage{amsmath,amsfonts,bbm,mathrsfs,amssymb}
\usepackage{slashed,cancel}
\usepackage[usenames,dvipsnames]{xcolor}
\usepackage{cite}
\usepackage{bm} 
\usepackage{hyperref}
\hypersetup{colorlinks=true,urlcolor=blue,anchorcolor=blue,citecolor=blue,filecolor=blue,
            linkcolor=blue,menucolor=blue, linktocpage=true,pdfproducer=medialab}
\usepackage[english]{babel}
\usepackage{catchfile}
\usepackage{indentfirst}
\usepackage[scriptsize]{subfigure}
\usepackage{graphicx}
\usepackage{float}
\usepackage{enumerate}
\usepackage{multirow}
\usepackage{tikz-feynman}
\usepackage[normalem]{ulem} 
\usepackage{ytableau}
\usepackage{xcolor}
\usepackage{fancyvrb}
\usepackage{makecell}
\usepackage{amsmath}
\usepackage{orcidlink}

\newcommand{\email}[1]{\href{mailto:#1}{\tt #1}}

\numberwithin{equation}{section}

\newcommand{\be}{\begin{equation}}
\newcommand{\ee}{\end{equation}}
\newcommand{\ba} {\begin{equation}\begin{aligned}}
\newcommand{\ea} {\end{aligned}\end{equation}}
\newcommand{\bg} {\begin{equation}\begin{gathered}}
\newcommand{\eg} {\end{gathered}\end{equation}}

\newcommand{\sL}{\mathscr{L}}

\newcommand{\mueV}{\ \mu\text{eV}}

\newcommand{\GeV}{\ \text{GeV}}

\newcommand{\cm}{\ \text{cm}}

\newcommand{\NSCool}{\texttt{NSCool}~\cite{NSCool}}

\newcommand{\bea}{\begin{eqnarray}}
\newcommand{\eea}{\end{eqnarray}}

\begin{document}
	
\vspace{2cm}

\begin{center}
	\boldmath
	
	{\textbf{\LARGE Core Composition Effects on the QCD Axion Mass Limit from Neutron Star Cooling}}
	\renewcommand{\thefootnote}{\fnsymbol{footnote}}
	\unboldmath
	
	\bigskip
	
	\vspace{0.5 truecm}
	
	{\bf Fernando Arias-Arag\'on\orcidlink{0000-0002-1161-840X}}\footnote{\email{fernando.arias@urjc.es}}$^{,a}$ and {\bf Federico Nola\orcidlink{0000-0001-9249-9547}}\footnote{\email{federico.nola@unicampania.it}}$^{,b,c,d}$\\
    {$^a$\it Departamento de Geología, Física y Química Inorgánica, Universidad Rey Juan Carlos, Calle Tulipán s/n, 28933 Móstoles, Madrid, Spain}\\
	{$^b$\it Dipartimento di Matematica e Fisica, Università degli Studi della Campania “Luigi Vanvitelli”, viale Abramo Lincoln 5- I-81100 Caserta, Italy}\\
	{$^c$\it Istituto Nazionale di Fisica Nucleare, Sezione di Napoli, Strada Comunale Cinthia, 80126 Napoli NA}\\
    {$^d$\it Istituto Nazionale di Fisica Nucleare, Laboratori Nazionali di Frascati, C.P. 13, 00044 Frascati, Italy}\\
	\vspace{2cm}

    \setcounter{footnote}{0}
\renewcommand{\thefootnote}{\arabic{footnote}}
	{\bf Abstract }
\end{center}

    Neutron stars are very dense media in which axions may be produced. This has been used to set a limit on the QCD axion mass, usually under the assumption that only neutrons, protons, electrons, and muons appear in the star core. Given the extreme conditions reached within neutron stars, it is reasonable to consider that other particles, such as hyperons and $\Delta$ resonances, may appear in the neutron star core. In this work, we study how the limit on the mass of QCD axions, namely KSVZ and DFSZ invisible axions, is altered when different equations of state are used, allowing for heavier particles to appear in the neutron star core. We find that this dependence is in general mild and thus reinforces the reliability of the known limit. Additionally, in the DFSZ scenario, it may drive the limit within the sensitivity window for IAXO. This would allow this experiment to discern the composition of neutron star cores if an axion were to be observed within that window.

\thispagestyle{empty}
\vfill

\newpage

\tableofcontents
\vspace{-0.25cm}

\section{Introduction} \label{sec:Intro}

Neutron Stars (NSs) represent an ideal environment for testing high energy physics phenomena due to the high densities and pressures reached in their cores~\cite{Lattimer:2004sa}. Their thermal evolution is largely dependent on the emission of very weakly coupled particles, such as neutrinos ~\cite{Page:1997mj,Yakovlev:2000jp}, that escape the dense core and allow the star to lose energy very efficiently. The emission rate of these NS neutrinos will thus depend on the components one assumes for the NS crust and core; particularly, the core composition of a NS is a topic still open to discussion~\cite{Page:2005fq}, with possibilities ranging from neutron matter, with free protons, neutrons, and leptons, to more exotic possibilities~\cite{Maxwell:1986pj,Prakash:1992zng,Muto:1994unh} such as quark-gluon plasma~\cite{Ivanenko:1969bs,Iwamoto:1982zz}. All this information, encoded in the Equation of State (EOS), affects not only the emission of neutrinos and other particles throughout the life of the NS but also its formation and stability~\cite{Lattimer:2004pg,Yakovlev_2004,Page:2005fq,Cruz-Camacho:2024odu}. In particular, different core compositions can open or suppress specific cooling channels. The appearance of hyperons, $\Delta$ resonances, or other exotic degrees of freedom may modify the behaviour of the cooling processes. Therefore, changing the microscopic composition of the core can alter the standard neutrino cooling history.

A thorough analysis of any beyond-the-Standard-Model (BSM) observable would require the study of such uncertainties. A prototypical example of a BSM particle often mentioned in the context of NSs is the quantum chromodynamics (QCD) axion~\cite{Peccei:1977hh,Peccei:1977ur,Wilczek:1977pj,Weinberg:1977ma}. The QCD axion is a pseudoscalar\footnote{Limits on other BSM particles can also be found in the literature~\cite{Fiorillo:2025zzx,Fiorillo:2026wso,Bramante:2023djs,Dey:2025atz,Hao:2025fea}.} that arose as an explanation for the absence of CP violation in the strong sector~\cite{Abel:2020pzs}, and later on it was discovered to be an excellent dark matter  candidate~\cite{Preskill:1982cy,Abbott:1982af,Dine:1982ah}. Its mass is required to be small due to several constraints (among which supernovae (SN)~\cite{Carenza:2019pxu} and red giant observations~\cite{Viaux:2013lha,Straniero:2018fbv}), it interacts feebly with standard model (SM) particles, and would thus represent an excellent coolant for NSs if produced within them~\cite{Leinson:2014ioa,Hamaguchi:2018oqw,Sedrakian:2015krq,Carenza:2024ehj,Keller:2012yr,Sedrakian:2018kdm,Yadav:2025tmq,Gomez-Banon:2024oux,Fiorillo:2025gnd,Fiorillo:2026dqu,Kachelriess:1997kn,Leinson:2019cqv,Beznogov:2018fda}. In the extremely dense interior of a neutron star, axions can be produced through nucleon and baryon bremsstrahlung processes~\cite{Iwamoto:1984ir,Iwamoto:1992jp}. If the relevant baryonic species are paired in superfluid states, axion emission can also arise from Cooper pair breaking and formation (PBF)~\cite{Keller:2012yr}, in complete analogy with the corresponding neutrino processes~\cite{Yakovlev:1998wr}. These mechanisms make axion cooling directly sensitive to the core composition and to the assumed superfluidity model. NS cooling limits currently yield the strongest bound on the axion-neutron coupling constant $g_{ann}$~\cite{ParticleDataGroup:2024cfk,Buschmann:2021juv}, which can then be translated into a limit on the axion mass $m_a$ in well-defined scenarios like the Kim-Shifman-Vainshtein-Zakharov (KSVZ)~\cite{Kim:1979if,Shifman:1979if} and Dine-Fischler-Srednicki-Zhitnitsky (DFSZ)~\cite{Dine:1981rt,Zhitnitsky:1980tq} invisible axion frameworks. While the dependence of this limit on many uncertainties, such as EOS, neutron star mass, envelope, and superfluidity modelling in the core, were taken into account,  previous analyses were mostly performed assuming a single composition for the NS core: neutrons, protons, electrons, and muons\footnote{In the context of SNe and proto-neutron stars there has been some progress beyond this (cf.~\cite{Cavan-Piton:2024ayu,Camalich:2020wac}).}. The impact of varying the particle content of the core on axion cooling bounds has therefore remained rather unexplored. In the present analysis, the axion emission sector is restricted to nucleonic bremsstrahlung and nucleonic PBF processes. Although the QCD axion can also couple non-negligibly to hyperons~\cite{Vonk:2021sit}, calculating axion emissivities involving hyperons or $\Delta$ resonances is beyond the scope of this paper, and thus it is not included. These non-nucleonic degrees of freedom nevertheless affect the thermal evolution through the EOS, stellar structure, particle fractions, heat capacities, and neutrino emission channels. The resulting axion limits should therefore be interpreted within the nucleonic axion emission framework adopted in this work.

Our aim in this article is to build upon the previous analysis and study the universality of the limit that NS can set on axion physics. In order to do so, we will make use of the MUSES Calculation Engine (\texttt{MUSES CE})~\cite{MusesCalculationEngine, ReinkePelicer:2025vuh}, which allows the construction of NS EOSs with varying particle composition in the core. We supplement these models with EOSs from the DS~(CMF) family~\cite{Dexheimer:2008ax,Dexheimer:2009hi,Dexheimer:2017nse,Dexheimer:2020rlp},  considering analogous microscopic core compositions. These EOSs are used as input in the widely-used code \NSCool, where the corresponding neutrino and axion emissivities are implemented consistently with the assumed particle content and superfluid pairing model~\cite{1994AstL...20...43L}. We will confront the mass-radius curve obtained for each EOS with current bounds on NSs~\cite{Miller:2021qha} and, for those allowed, we will include axions and study the limits resulting from different core compositions, following the same procedure as was done in Ref.~\cite{Buschmann:2021juv} for neutron matter. We will perform the analysis for both benchmark classes of QCD axion models, namely the KSVZ and DFSZ models. This enables us to derive updated constraints on axion cooling and to assess the robustness of the corresponding $m_a$ limits against changes in the NS core composition. Our main goal is to determine whether bounds obtained under the standard nucleonic assumption remain stable once more general EOSs, including particles beyond neutrons, protons and leptons, are taken into account.

This article is organized as follows. In Sec.~\ref{sec:Models}, we briefly review the aspects of axion physics relevant to neutron star cooling and introduce the KSVZ and DFSZ benchmark models considered in our analysis. In Sec.~\ref{sec:EOS} we summarize how the open code \texttt{MUSES CE} ~\cite{MusesCalculationEngine} builds EOS for NS. Sec.~\ref{sec:CompoProcess} contains the description of the four scenarios we consider in this work, varying the particle content in the NS core, and gives a detailed list of the processes we added to neutrino cooling. Then, in Sec.~\ref{sec:SeqCool} we check the consistency of our scenarios with current observations. Finally, Sec.~\ref{sec:Results} presents our new results for the limits on the axion mass, both for KSVZ and DFSZ models, and in Sec.~\ref{sec:Conclusions} we close this work with some final remarks.

\section{Axion emission and models}\label{sec:Models}
\label{sec:QCDAxion}

Neutron stars cool primarily through neutrino emission from their interiors and photon emission from their surfaces. If axions exist, their production in dense stellar matter would provide an additional energy loss channel, as they can escape from the star and accelerate its thermal evolution. In this section, we summarize the nucleonic axion emission processes included in our cooling simulations and introduce the two benchmark axion models used to derive the corresponding mass limits.

In general, the axion~\cite{Peccei:1977hh,Peccei:1977ur,Weinberg:1977ma,Wilczek:1977pj} $a$ is a neutral pseudoscalar particle, usually light, which arises as pseudo Nambu-Goldstone boson after an anomalous global symmetry, the Peccei-Quinn (PQ) symmetry, is broken spontaneously. These particles may develop a coupling to any SM nucleon $N$, $c_{aN}$, which can be parametrized as shown in the following Lagrangian:

\be
\sL_a\supset \frac{c_{aN}}{2f_a} \bar{\psi}_N \gamma^\mu\gamma_5 \psi_N \partial^\mu a,
\ee
where $\gamma^\mu$ and $\gamma_5$ are the usual Dirac matrices, $\psi_N$ represents the nucleon fields with $N=n,p$ and $f_a$ is the characteristic energy scale of the axion, related to its mass $m_a$ by the following equation~\cite{GrillidiCortona:2015jxo}:
\be\label{eq:ma}
m_a\simeq 5.7\mueV \times\frac{10^{12}\GeV}{f_a}.
\ee

From this, it is common to redefine the coupling in a dimensionless fashion, $g_{aNN}$, such that:
\be\label{eq:gaNN}
g_{aNN}=\frac{c_{aN}\cdot m_N}{f_a},
\ee
where $m_N$ represents the nucleon mass.

Through this coupling, axions can be emitted in processes analogous to those where a pair of neutrino and antineutrino is involved, namely the bremsstrahlung processes 
\be
\begin{aligned}
    n+n &\to n +n+a,\\
    p+p &\to p +p+a,\\
    n+p &\to n +p+a,\\
\end{aligned}
\ee
and the nucleon pair-breaking formation process
\be
\begin{aligned}
    \{NN\} \leftrightarrow N+N +a.
\end{aligned}
\ee

For the axion emission processes considered in this work, we adopt the emissivities reported in the supplementary material of Ref.~\cite{Buschmann:2021juv}. These expressions include the relevant in-medium effects and provide the microphysical input for the cooling simulations described in Sec.~\ref{sec:CompoProcess}.

The axion emission sector employed in the present calculations is restricted to the nucleonic processes listed above. Direct axion emission channels involving hyperons or $\Delta$ resonances are not included. This should not be interpreted as implying that the corresponding emissivities vanish: the QCD axion can couple non-negligibly to the baryon octet \cite{Vonk:2021sit}. Their consistent implementation would require a dedicated analysis that lies beyond the scope of this work, and therefore the axion luminosities presented here must be interpreted in the context of purely nucleonic emission.

The axion luminosity depends on the axion mass and its couplings to nucleons. We focus on two phenomenologically relevant benchmark models: the KSVZ axion~\cite{Kim:1979if,Shifman:1979if} and the DFSZ axion~\cite{Zhitnitsky:1980tq,Dine:1981rt}. A comprehensive review of their properties can be found in Ref.~\cite{DiLuzio:2020wdo}; here, we summarize only the features relevant to the axion--nucleon couplings employed in our analysis.

In the case of KSVZ models, SM particles are neutral under the PQ symmetry, and thus the axion does not couple to SM quarks at tree level. It does, however, couple to gluons, as required in order to solve the strong CP problem, and this translates into an axion-proton and axion-neutron coupling, $c_{ap}$ and $c_{an}$ respectively~\cite{DiLuzio:2020wdo,GrillidiCortona:2015jxo}:
\ba\label{eq:KSVZ}
c^{KSVZ}_{ap} = -0.47\pm 0.03,\\ c^{KSVZ}_{an}=-0.02\pm 0.03.
\ea

On the other hand, DFSZ axions do couple at tree-level to SM quarks. This class of models requires the presence of two Higgs doublets, and therefore introduces the parameter $\tan\beta=v_u/v_d$, which is simply the ratio of the vacuum expectation value of the up-type Higgs doublet divided by that of the down-type. With this, one can find that the DFSZ axion coupling to nucleons is written as~\cite{DiLuzio:2020wdo,GrillidiCortona:2015jxo}:

\ba\label{eq:DFSZ}
c^{DFSZ}_{ap} = -0.182\pm 0.025-0.435\sin^2\beta,\\ c^{DFSZ}_{an}=-0.160\pm0.025+0.414\sin^2\beta.
\ea

\section    {Construction of the NS EOS with the \texttt{MUSES CE}} \label{sec:EOS}
\begin{figure}
    \centering
    \includegraphics[width=0.5\linewidth]{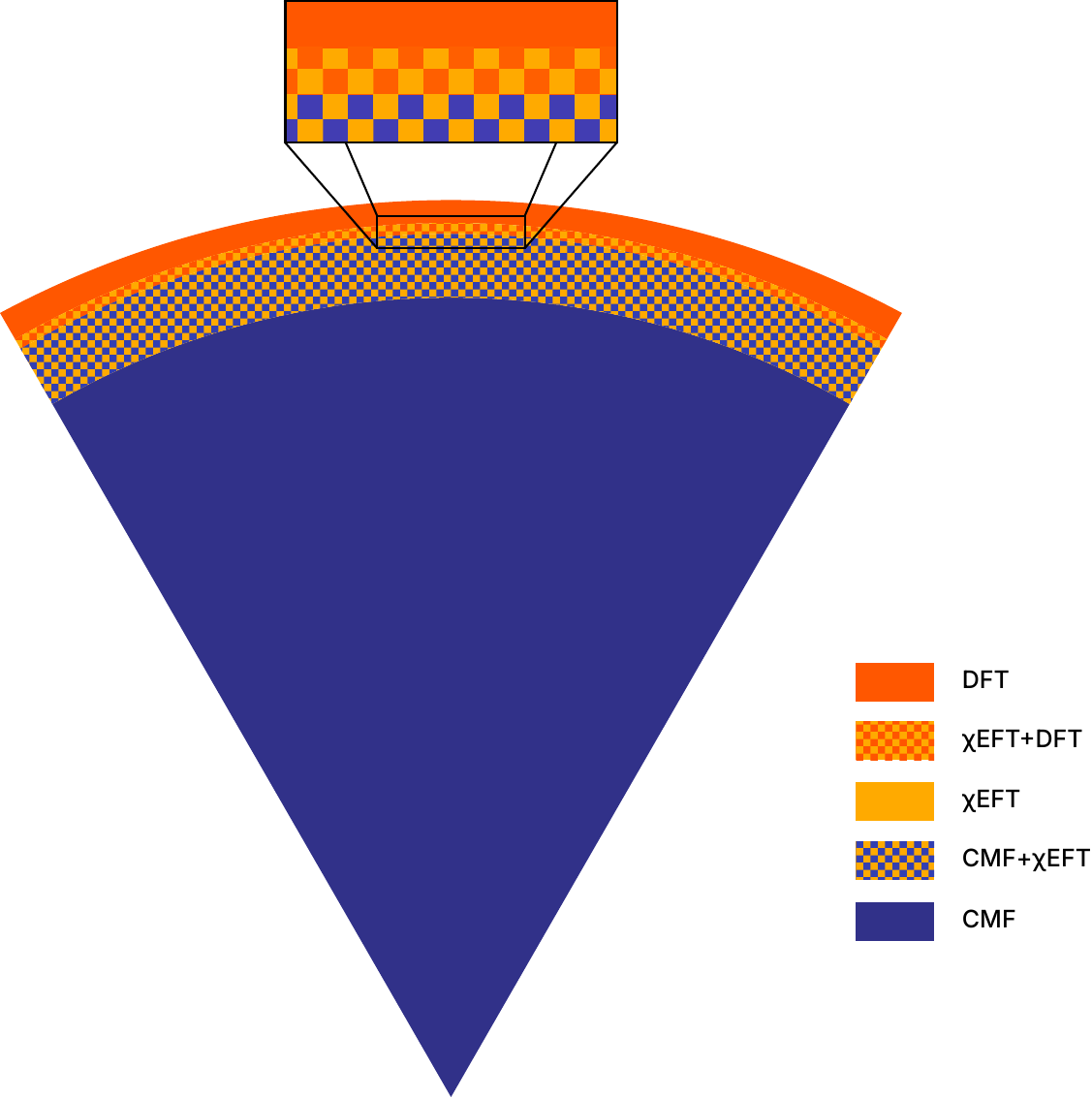}
    \caption{Visual example of neutron star slice that can be obtained with \texttt{MUSES CE}. The different colours indicate the different models used, as shown in the legend. The proportions of the layers depend on the mass of the star.}
    \label{fig:NS_slice}
\end{figure}

The construction of the equations of state was done through the MUSES Calculation Engine, where MUSES stands for “Modular Unified Solver of the Equation of State” \cite{MusesCalculationEngine, ReinkePelicer:2025vuh}. The strength of this approach in constructing the EOS of a neutron star is the possibility of using a different theoretical approach for the different density regions of the star, as shown in Fig. \ref{fig:NS_slice}. \texttt{MUSES CE} can combine a crust calculation, an intermediate-density nuclear-matter calculation, a supranuclear core model, together with the leptonic sector needed for charge neutrality and $\beta$-equilibrium and a synthesis stage that matches the different density windows, to build a complete EOS for NS. In compact form, the neutron star construction workflow may be represented schematically as:

{\footnotesize
\begin{equation*}
\mathcal{W}_{\rm NS}:
\bigl\{\mathrm{Crust\!-\!DFT},\chi\mathrm{EFT},\mathrm{CMF}\bigr\}
\xrightarrow{\ \mathrm{Lepton}\ }\bigl\{\beta\text{-equilibrated one-dimensional EOS}\bigr\}
\xrightarrow{\ \mathrm{Synthesis}\ }
\varepsilon(P).
\end{equation*}}

This factorization is not only computationally convenient. It reflects the fact that the outer crust, the outer core and the inner core are controlled by different effective degrees of freedom and constraints.

\subsection{Low density regions and around saturation density}\label{sec:Crust}
At low baryon densities, the module employed is based on Crust Density Functional Theory (Crust-DFT) \cite{Negele:1971vb,Chamel:2008ca}. This module provides a free-energy density description of matter in the thermodynamic space $(n_B,Y_e,T)$. Its construction allows for a consistent interpolation among different physical regimes, including nuclear matter close to saturation constrained by phenomenology, cold neutron-rich matter constrained by microscopic calculations, low density hot matter and the inhomogeneous phase where nuclei coexist with a gas of dripped nucleons \cite{Du2019,Du2022}.

A key feature of the module is the explicit separation of the rest-mass contribution from the free-energy density. Denoting by $\xi$ the volume fraction available to the nucleon gas, by $n_n'$ and $n_p'$ the local neutron and proton densities outside nuclei, and by $n_i$ the density of nuclei with neutron and proton numbers $(N_i,Z_i)$, one has
\begin{align}
 f_{\mathrm{rest}}
 &=
 \xi n_n^{\prime} m_n
 + \xi n_p^{\prime} m_p
 + \sum_i N_i n_i m_n
 + \sum_i Z_i n_i m_p  \nonumber \\
 &=
 n_B(1-Y_e)m_n + n_B Y_e m_p .
\end{align}

This identity makes the conversion between internal and total thermodynamic quantities transparent, which is essential when the crust equation of state is matched to homogeneous matter.

The module also keeps track of the transformation between different composition variables. Matching conditions and stability analyses may be formulated in terms of $(n_n,n_p)$, $(n_B,n_e)$, or $(n_B,Y_e)$. Therefore, the relevant derivatives are transformed exactly, for example
\begin{align}
\left(\frac{\partial}{\partial n_B}\right)_{Y_e}
=
(1-Y_e)
\left(\frac{\partial}{\partial n_n}\right)_{n_p}
+
Y_e
\left(\frac{\partial}{\partial n_p}\right)_{n_n},
\end{align}
and
\begin{align}
\left(\frac{\partial}{\partial Y_e}\right)_{n_B}
=
n_B
\left[
\left(\frac{\partial}{\partial n_p}\right)_{n_n}
-
\left(\frac{\partial}{\partial n_n}\right)_{n_p}
\right],
\end{align}
with analogous relations when additional leptonic degrees of freedom, such as muons, are included \cite{Du2022,ReinkePelicer:2025vuh}.

\begin{figure}[t!]
    \centering
    \includegraphics[width=0.6\linewidth,trim=0.25 16 0.5 0.75,clip]{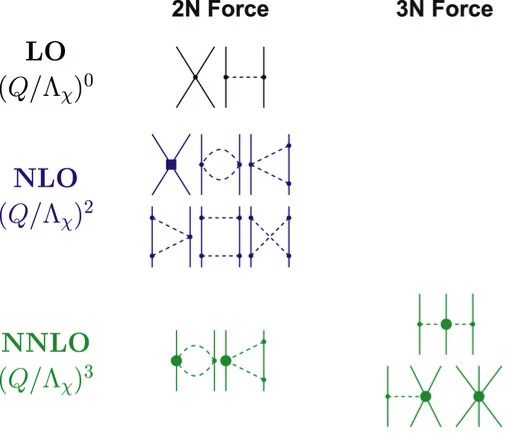}
    \caption{Hierarchy of chiral nuclear
interactions as a function of chiral order, showing the emergence of the two-nucleon and three-nucleon forces.}
    \label{fig:hierarchy}
\end{figure}

As a bridge between low density regions and high density regions, around nuclear saturation density, a Chiral Effective Field theory ($\chi$EFT)-based module is employed \cite{Machleidt:2011zz}. The starting point is the most general effective Lagrangian, consistent with the symmetries of low energy QCD, 
\begin{align}
\mathcal{L}_{\rm eff}
=
\mathcal{L}_{\pi\pi}
+
\mathcal{L}_{\pi N}
+
\mathcal{L}_{NN}
+
\mathcal{L}_{3N}
+
\cdots .
\end{align}
The resulting nuclear interaction is ordered according to the Weinberg power counting \cite{Weinberg1990}, i.e, as an expansion in powers of the soft scale $Q$ over the chiral breakdown scale $\Lambda_\chi$.
This generates a controlled hierarchy of many-body forces, where the first non-vanishing three nucleon force term appears at NNLO, as shown in Fig.~\ref{fig:hierarchy}. In the context of the \texttt{MUSES CE}, the $\chi$EFT EOS is evaluated through a many-body perturbation theory \cite{Wellenhofer:2014hya}, employing the chiral nuclear interaction with cut-off $\Lambda=450$ MeV \cite{Coraggio:2012ca}.

\subsection{High density regions}\label{sec:Core}
In the deep core region, with higher density, the description is performed by the Chiral Mean Field (CMF) module. The CMF framework is a relativistic mean field realization of a non linear $SU(3)$ sigma model, in which hadrons interact via scalar and vector meson fields, and most effective masses are generated dynamically through their couplings to the surrounding medium \cite{Papazoglou:1998vr,Dexheimer:2008ax,Dexheimer:2009hi}. Schematically, the CMF Lagrangian can be written as
\begin{align}
\mathcal{L}_{\rm CMF}
=
\mathcal{L}_{\rm kin}
+
\mathcal{L}_{\rm int}
+
\mathcal{L}_{\rm self}
+
\mathcal{L}_{\rm SB}
-
U(\Phi),
\end{align}
where $\mathcal{L}_{\rm kin}$ contains the kinetic terms,
$\mathcal{L}_{\rm self}$ encodes the mesonic self interactions, $\mathcal{L}_{\rm SB}$ accounts for explicit symmetry breaking effects and $U(\Phi)$ describes the coupling to the Polyakov-loop variable. The interaction sector has the form
\begin{align}
\mathcal{L}_{\rm int}
=
-\sum_i \bar{\psi}_i
\left[
\gamma_0
\left(
g_{i\omega}\omega
+
g_{i\phi}\phi
+
g_{i\rho}\tau_3\rho
\right)
+
M_i^*
\right]
\psi_i ,
\end{align}
with effective baryon masses generated by the scalar condensates,
\begin{align}
M_i^*
=
g_{i\sigma}\sigma
+
g_{i\delta}\tau_3\delta
+
g_{i\zeta}\zeta
+
\delta m_i .
\end{align}

Thus, changes in the scalar mean fields modify the in-medium particle masses, while the vector fields shift the corresponding chemical potentials.

In the modern CMF++ implementation, the thermodynamics is naturally formulated in the three-dimensional charge space spanned by baryon number, electric charge and strangeness. The chemical potential \cite{Aryal:2020ocm} of each species is written as
\begin{align}
\mu_i
=
B_i\mu_B
+
Q_i\mu_Q
+
S_i\mu_S ,
\end{align}
while the effective chemical potential entering the single-particle spectrum is
\begin{align}
\mu_i^*
=
\mu_i
-
g_{i\omega}\omega
-
g_{i\phi}\phi
-
g_{i\rho}\tau_3\rho .
\end{align}

This formulation allows to obtain pressure as a function of $(\mu_B,\mu_Q,\mu_S)$, with the physical branch selected by the appropriate phase-stability conditions \cite{Cruz-Camacho:2024odu}.

The main relevance of the recent CMF++ reformulation is not only computational. The crucial conceptual advantage is that stable, metastable and unstable branches can be mapped in the full chemical potential space. This also enables the calculation of higher-order susceptibilities associated with the deconfinement transition
\cite{Cruz-Camacho:2024odu}. In particular, the local phase structure is encoded in the curvature matrix of the pressure,
\begin{align}
\chi_{ab}
=
\frac{\partial^2 P}
{\partial \mu_a \partial \mu_b},
\qquad
a,b\in\{B,Q,S\}.
\end{align}

The complete set of eigenvalues of $\chi_{ab}$ distinguishes mechanically and chemically stable regions from metastable or unstable branches. For the NS applications considered here, the general CMF++ framework is restricted to the cold, charge-neutral and $\beta$-equilibrated trajectory, using the C4 coupling set. 

The inclusion of hyperons generally softens the high density EOS and may reduce the maximum neutron star mass, giving rise to the so-called ``hyperon puzzle"~\cite{Takatsuka:2004ch,Lonardoni:2014bwa,Bombaci:2016xzl,Vidana:2019zuz}. In the CMF parametrizations considered here, however, the resulting EOSs remain sufficiently stiff to support neutron stars with masses of approximately $2\,M_\odot$ \cite{ReinkePelicer:2025vuh}. This result depends on the adopted interaction model and should not be regarded as a general resolution of the hyperon puzzle, which has been addressed in the literature~\cite{Weissenborn:2011ut,Astashenok:2014pua,Katayama:2015dga,Raduta:2017wpp,Vidana:2022tlx}.

\subsection{Lepton sector \& Synthesis}\label{sec:Synthesis}
A purely baryonic EOS is not sufficient to fully describe a NS, so the addition of a leptonic sector is crucial, ensuring charge neutrality and $\beta$-stability \cite{ReinkePelicer:2025vuh}. In this sense, the Lepton module is the component that converts a baryonic EOS into the physically relevant composition for cold neutron star matter.

The module starts from the free-fermion Lagrangian \cite{Hanauske:1999ga}
\begin{align}
\mathcal{L}
=
\sum_{i=e,\mu,\tau}
\left[
\bar{\psi}_i
\left(
i\gamma_\mu\partial^\mu
-
m_i
\right)
\psi_i
+
\bar{\nu}_{Li}
i\gamma_\mu\partial^\mu
\nu_{Li}
\right],
\end{align}
which leads, at $T=0$, to analytic expressions for the thermodynamic quantities of each lepton species. For a charged lepton $\ell$, one has
\begin{align}
n_\ell
=
\frac{\gamma_\ell}{6\pi^2} k_{F,\ell}^3 ,
\qquad
\mu_\ell
=
\sqrt{k_{F,\ell}^2+m_\ell^2},
\end{align}
where $k_{F,\ell}$ is the Fermi momentum, $m_\ell$ is the lepton mass, and $\gamma_\ell$ is the spin degeneracy factor. Once we have the corresponding energy density, $\varepsilon_\ell$, and pressure, $P_\ell$, all these quantities are tabulated by the module and subsequently combined with the baryonic contribution to build the total thermodynamics. For cold NS matter, the equilibrium and neutrality conditions are 
\begin{align}
n_Q- \sum_{\ell}n_{\ell}=0, \qquad -\mu_Q=\mu_e,
\end{align}
since in \texttt{MUSES CE} formalism different flavours of leptons have the same chemical potential, $\mu_e=\mu_\mu=\mu_\tau$, because they have the same quantum numbers.

Once the baryonic EOSs have been completed with the leptonic sector, in $\beta$-equilibrium and charge neutrality, the Synthesis module is used to combine them together. It is important to remark that this module does not simply paste the various number tables one after the other, but implements several physically distinct prescriptions for joining the different equations of state. For the purpose of this paper, the matching between the different density sectors was done using a hyperbolic tangent interpolation, which provides a smooth transition between two EOSs over a finite interval of chosen thermodynamic variable $x$ \cite{ReinkePelicer:2025vuh}. In this approach \cite{Masuda:2012ed}, a thermodynamic quantity $Y$ is written as
\bg
Y(x)
=
Y^{(I)}(x) f_-(x)
+
Y^{(II)}(x) f_+(x),
\\
f_+(x)+f_-(x)=1 ,
\eg
where the switching functions are
\begin{align}
f_\pm^{\tanh}(x)
=
\frac{1}{2}
\left[
1
\pm
\tanh\left(
\frac{x-\bar{x}}{\Gamma}
\right)
\right].
\end{align}
Here $\bar{x}$ denotes the centre of the matching region, while $\Gamma$ controls the width of the transition. 

\section{Core compositions and neutrino emission}\label{sec:CompoProcess}

In this section, we discuss the NS core compositions considered in this work. Depending on the particles allowed to be present within the NS core, the processes available for neutrino emission will differ, and thus the thermal history of the NS may change. Neutrino emission processes can be broadly divided into four categories: direct and modified URCA~\cite{1995A&A...297..717Y} (DURCA and MURCA respectively henceforth) processes, bremsstrahlung processes, and pair breaking and formation (PBF) processes when superfluidity is considered. When particles beyond those of neutron matter are considered, a large variety of reactions show up that can alter the neutrino emission and lead to a potentially different cooling. Some of these processes, while not considered in most previous works, were already ready to be used within the public version of the \texttt{NSCool} evolution code; some others, however, we include here for the first time, upgrading the cooling code to a more powerful, versatile, and robust version.

\subsection{$npe\mu$}\label{sec:npemu}

Most previous studies of NS cooling limits on axion couplings considered that the core within a NS is composed solely of neutron matter. Although the crust can be described with a crystalline structure of nuclei and free electrons, at the higher densities reached within the core these nuclei break down into neutrons and protons, and high degeneracy may be achieved. Therefore, one can describe the core of NS as a homogeneous dense matter made of neutrons, protons, electrons, and, when energetically allowed, muons.

In this case, the relevant neutrino emission processes are:
\begin{itemize}
    \item Direct URCA: $n\xrightarrow{}p+e+\bar{\nu}_e$ and  $p+e\xrightarrow{}n+\nu_e$
    \item Modified URCA: $N+n\xrightarrow{}N+p+e+\bar{\nu}_e$ and  $N+p+e\xrightarrow{}N+n+\nu_e$, with $N=n,p$
    \item Bremsstrahlung: $N+N^\prime\rightarrow N+N^\prime+\nu+\bar{\nu}$, with $N,N^\prime=n,p$
    \item PBF in the singlet gaps $^1S_0$ for neutrons and protons
\end{itemize}

The emissivities for all these processes were already included in the code \NSCool $\ $ with expressions matching those of the seminal work of Ref.~\cite{Friman:1979ecl,Prakash:1992zng,Yakovlev:1998wr}. In our work, however, we modify them in two ways: on the one hand, we use for MURCA and bremsstrahlung processes the emissivities computed in \cite{Bottaro:2024ugp}, which include rho meson exchanges and do not use the triangular approximation but do not consider in-medium effects. On the other hand, we include said effects by multiplying those emissivities from~\cite{Bottaro:2024ugp} by the factor $\gamma^6 (m_N^*/m_N)^{-4}$ as prescribed in the supplementary material of Ref.~\cite{Buschmann:2021juv}, from where we also use the emissivities for DURCA and PBF processes.

\begin{figure}[t!]
    \centering
    \includegraphics[width=0.7 \textwidth,trim=1pt 1pt 1pt 1pt,clip]{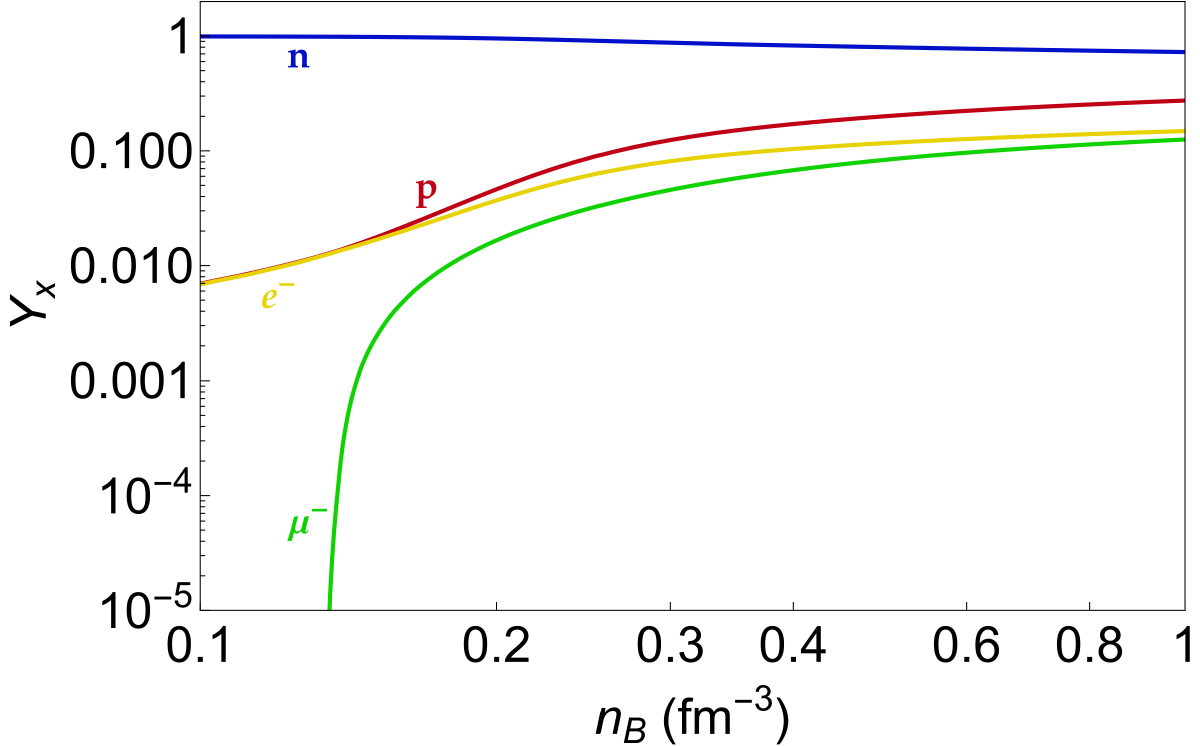}
    \caption{ Particle fractions within the NS core for the different species allowed when using the \texttt{MUSES CE} to produce an EOS in the \textit{$npe\mu$} case.}
    \label{fig:npemuCompo}
\end{figure}

In Fig.~\ref{fig:npemuCompo} we show the particle fraction $Y_X$ in the core of the NS
\be
Y_X = \frac{n_X}{n_B},
\ee
where $n_X$ and $n_B$ represent the number density of particle $X$ and the total baryon number density, respectively. We build the EOS with the \texttt{MUSES CE} and label this the case \textit{$npe\mu$}. Charge neutrality is, of course, ensured throughout all values of the baryon density. This represents the composition most commonly used so far in axion studies, and is the one considered for benchmark EOS such as APR or the BSk family~\cite{Akmal:1998cf,Pearson:2018tkr}. Additionally, we use the DS~(CMF) 2 EOS from CompOSE~\cite{CompOSEweb,Oertel:2016bki,CompOSECoreTeam:2022ddl} as an analogue from a different family (with the exception that DS~(CMF) 2 does not include muons).

\subsection{Spin $\frac{3}{2}$ resonances}\label{sec:Delta}

Departing from the canonically considered case, we now allow for the existence of baryonic resonances with spin $\frac{3}{2}$, still restricting ourselves to first-generation particles. We label this the \textit{Delta} scenario, as it implies allowing the $\Delta^-$, $\Delta^0$, $\Delta^+$ and $\Delta^{++}$ baryons to be present in the NS core.These baryons do not follow weak channels in their decays, so one may be tempted to disregard this case. However, allowing for the presence of said particles may alter the number densities of the constituents of neutronic matter, thus potentially altering the NS thermal history.

\begin{figure*}[t!]
    \centering
    \includegraphics[width=0.7 \textwidth,trim=0.75 1 0.5 0.75,clip]{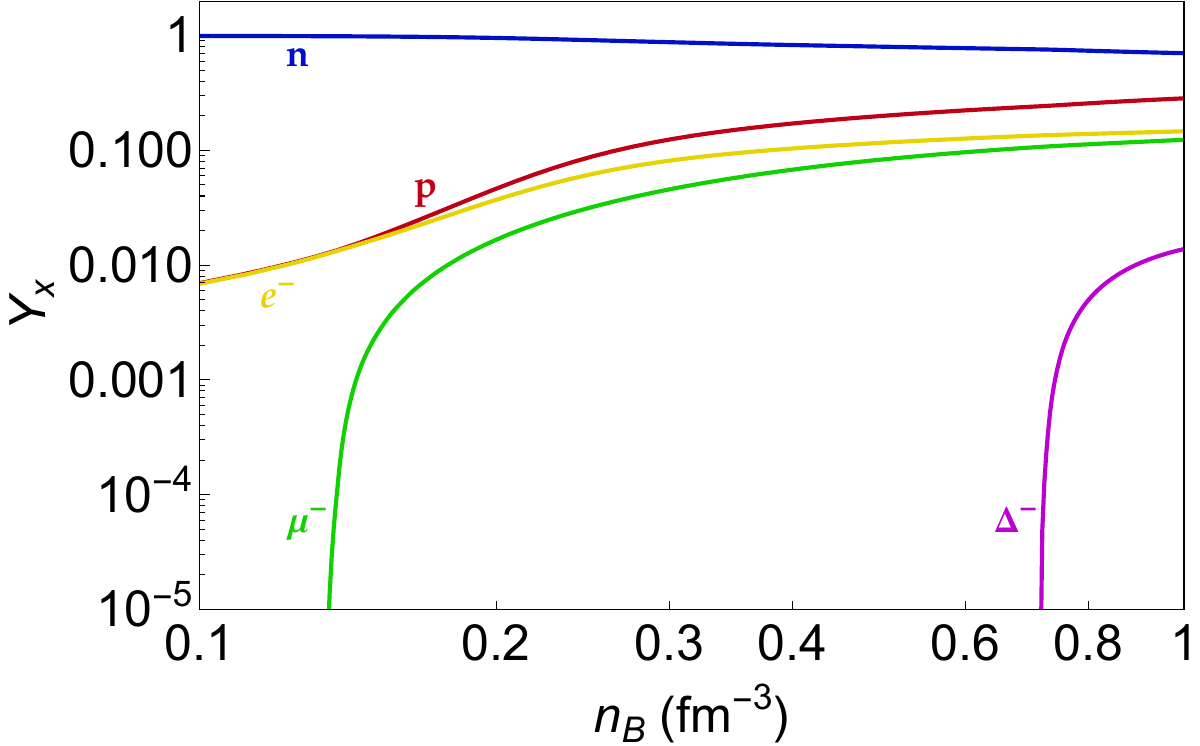}
    \caption{ Particle fractions within the NS core for the different species allowed when using the \texttt{MUSES CE} to produce an EOS in the \textit{Delta} case.}
    \label{fig:DeltaCompo}
\end{figure*}

In Fig.~\ref{fig:DeltaCompo} we show the particle content obtained in the \textit{Delta} scenario with the \texttt{MUSES CE}. Only deep within the NS core do these resonances show up, amounting to roughly $1\%$ of the composition at large baryon densities. The DS~(CMF) 8 EOS is analogous to the one obtained with the \texttt{MUSES CE} for this composition, except that it does not contain muons.

\subsection{Hyperons}\label{sec:Octet}

After including higher spin resonances, we now allow as well for baryons with strangeness. This implies considering the full baryonic octet and decuplet, including all hyperons. Most hyperons in the decuplet undergo strong decays to their spin $\frac{1}{2}$ partners from the octet, meaning that they do not need to be considered for neutrino emission. However, The inclusion of the full baryonic octet implies a large variety of weak processes to be considered. The stellar evolution code \texttt{NSCool} \cite{NSCool} did include some of them, namely the following DURCA processes:

\be
\begin{aligned}
& \Lambda \rightarrow p + \ell + \bar{\nu}_{\ell}, \quad
&&p + \ell \rightarrow \Lambda + \nu_{\ell}, \\
&\Sigma^- \rightarrow n + \ell + \bar{\nu}_{\ell}, \quad
&&n + \ell \rightarrow \Sigma^- + \nu_{\ell}, \\
&\Sigma^- \rightarrow \Lambda + \ell + \bar{\nu}_{\ell}, \quad
&&\Lambda + \ell \rightarrow \Sigma^- + \nu_{\ell}, \\
&\Sigma^- \rightarrow \Sigma^0 + \ell + \bar{\nu}_{\ell}, \quad
&&\Sigma^0 + \ell \rightarrow \Sigma^- + \nu_{\ell}.
\end{aligned}
\ee

In this work, we supplement this by including all other relevant processes for hyperonic neutrino emission within \NSCool. Starting with the remaining direct URCA processes~\cite{Prakash:1992zng}, we include:

\be
\begin{aligned}
&\Xi^- \rightarrow \Lambda + \ell + \bar{\nu}_{\ell}, \quad
&&\Lambda + \ell \rightarrow \Xi^- + \nu_{\ell} \\
&\Xi^- \rightarrow \Sigma^0 + \ell + \bar{\nu}_{\ell}, \quad
&&\Sigma^0 + \ell \rightarrow \Xi^- + \nu_{\ell} \\
&\Xi^0 \rightarrow \Sigma^+ + \ell + \bar{\nu}_{\ell}, \quad
&&\Sigma^+ + \ell \rightarrow \Xi^0 + \nu_{\ell} \\
&\Xi^- \rightarrow \Xi^0 + \ell + \bar{\nu}_{\ell}, \quad
&&\Xi^0 + \ell \rightarrow \Xi^- + \nu_{\ell}
\end{aligned}
\label{eq:hyp_durca}
\ee

Next, we also include the emissivities due to modified URCA processes where hyperons are involved~\cite{Maxwell:1986pj}:~
\be
\begin{aligned}
&\Sigma^- + \Sigma^- \rightarrow \Sigma^- + \Lambda + e^- + \bar{\nu}_e \\
&\Sigma^- + \Lambda \rightarrow \Lambda + \Lambda + e^- + \bar{\nu}_e \\
&\Sigma^- + n \rightarrow \Sigma^- + p + e^- + \bar{\nu}_e \\
&\Sigma^- + n \rightarrow \Lambda + n + e^- + \bar{\nu}_e \\
&\Sigma^- + p \rightarrow \Lambda + p + e^- + \bar{\nu}_e
\end{aligned}
\ee

\begin{figure}[t!]
    \centering
    \includegraphics[width=0.7 \textwidth,trim=0.75 1 0.5 0.75,clip]{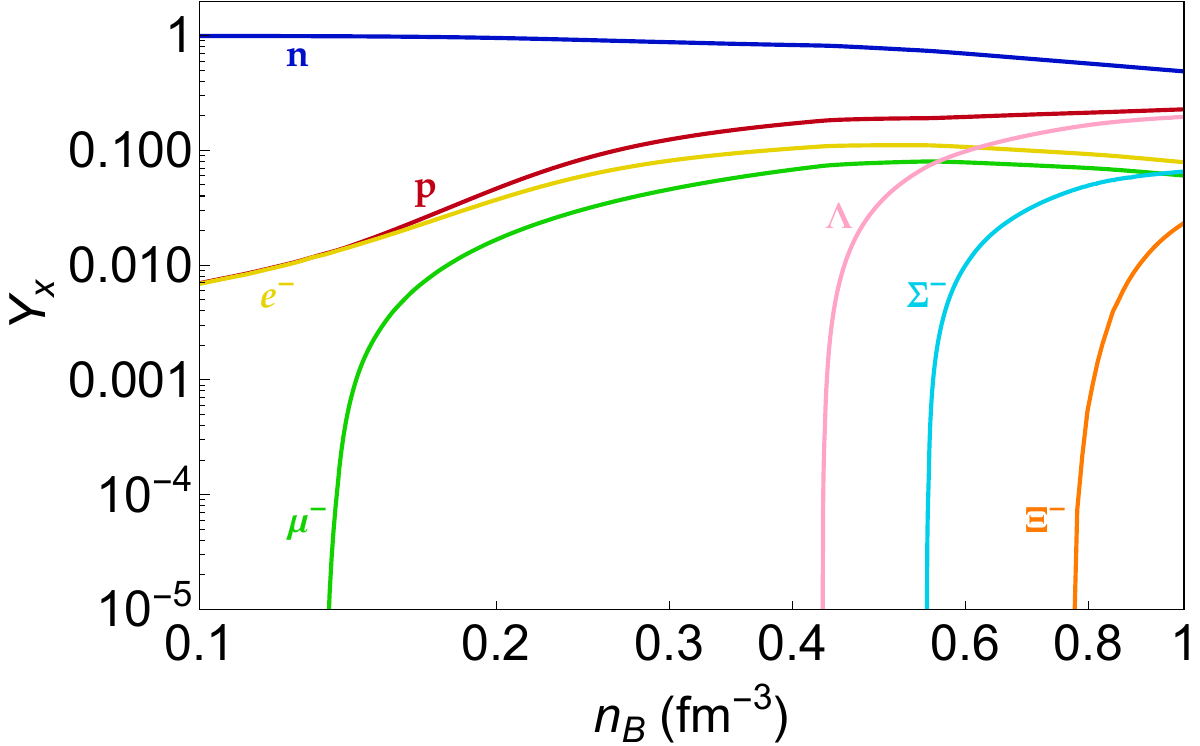}
    \caption{ Particle fractions within the NS core for the different species allowed when using the \texttt{MUSES CE} to produce an EOS in the \textit{Octet} case.}
    \label{fig:OctetCompo}
\end{figure}

Hyperonic bremsstrahlung can also be found within Ref.~\cite{Maxwell:1986pj}, and so we write their contribution down within \NSCool:

\be
\begin{aligned}
&\Sigma^- + n \rightarrow \Sigma^- + n + \nu + \bar{\nu} \\
&\Sigma^- + p \rightarrow \Sigma^- + p + \nu + \bar{\nu} \\
&\Sigma^- + \Lambda \rightarrow \Sigma^- + \Lambda + \nu + \bar{\nu} \\
&\Sigma^- + \Sigma^- \rightarrow \Sigma^- + \Sigma^- + \nu + \bar{\nu} \\
&\Sigma^- + p \rightarrow \Lambda + n + \nu + \bar{\nu}
\end{aligned}
\ee

The modified-Urca and bremsstrahlung reactions listed above do not include processes involving $\Xi$ hyperons. The $\Xi$-mediated direct Urca reactions listed in Eq.~(\ref{eq:hyp_durca}) are included, whereas the corresponding slow $\Xi$ channels are not implemented. When a $\Xi$ direct Urca reaction is kinematically allowed, its $T^6$ temperature dependence generally makes it dominant over modified Urca and bremsstrahlung processes, whose emissivities scale as $T^8$ \cite{Prakash:1992zng,Yakovlev:2000jp}. In addition, the $\Xi$ fractions remain comparatively smaller than the $\Lambda$ and $\Sigma$ ones in the stellar models considered here. We therefore expect the omitted slow channels to provide a subleading contribution within the present setup.

Finally, hyperons may also form Cooper pairs when the NS temperature approaches the critical temperature of each baryon. We assume that they only form pairs in the singlet $^1S_0$ state, and thus include their contribution to PBF processes~\cite{Yakovlev:1998wr}:

\be
\begin{aligned}
&\{\Sigma^- \Sigma^-\} \leftrightarrow \Sigma^- + \Sigma^- + \nu + \bar{\nu} \\
&\{\Sigma^+ \Sigma^+\} \leftrightarrow \Sigma^+ + \Sigma^+ + \nu + \bar{\nu} \\
&\{\Xi^0 \Xi^0\} \leftrightarrow \Xi^0 + \Xi^0 + \nu + \bar{\nu} \\
&\{\Xi^- \Xi^-\} \leftrightarrow \Xi^- + \Xi^- + \nu + \bar{\nu}
\end{aligned}
\ee

Allowing for the presence of hyperons leads to the last two scenarios considered: \textit{Octet} and \textit{Oct+Dec}. The first includes only spin $\frac{1}{2}$ baryons, while the second also includes the spin $\frac{3}{2}$ decuplet. Although the DS~(CMF) EOS 1 and 7 do present different compositions in these cases, the result offered by the \texttt{MUSES CE} is the same in both situations: when hyperons are included, no $\Delta$ resonance appears in the star, let alone heavier particles from the decuplet. This \texttt{MUSES CE} \textit{Octet} case is the one for which we display the particle fractions in Fig.~\ref{fig:OctetCompo}, which contrasts with the previous cases in the decrease of leptonic fraction at very high densities due to the presence of $\Sigma^-$ and $\Xi^-$ baryons.

\subsection{More Exotic Compositions}\label{sec:Exot}

When exploring core compositions beyond the standard one, a widely discussed possibility is the appearance of pion or kaon condensates. At sufficiently high densities, these bosonic degrees of freedom may form Bose--Einstein condensates, modifying the particle composition and opening additional neutrino emission channels. A summary of these processes, including estimates of the corresponding direct Urca emissivities, can be found in Ref.~\cite{Yakovlev:2000jp}; see also Refs.~\cite{Thorsson:1993bu,Campbell:1974qt,Ellis:1995kz}. Rapid neutrino emissivities in kaon condensates are considered in the following papers~\cite{Tatsumi:1988up,Brown:1988ik}.

The low-energy interactions of the QCD axion with the hadronic sector include axion--pion and axion--pion--nucleon couplings, which may provide additional axion production mechanisms in pion-rich matter. Pion-induced axion emission has been investigated in supernova and proto-NS contexts, where reactions such as $\pi^-+p\rightarrow n+a$ can significantly enhance the axion emissivity \cite{Choi:2021ign,Fischer:2021jfm}. Axion emission from cold NS matter containing pion or kaon condensates was considered in Ref.~\cite{Muto:1994unh}, where the presence of a meson condensate was found to potentially enhance axion cooling. That calculation was based on axion--nucleon interactions, and a modern treatment consistently including the complete set of relevant axion--meson--nucleon interactions would therefore be required. However, the current version of \texttt{MUSES CE} does not provide pion or kaon condensates and their abundances. We consequently do not include these compositions in the present analysis and leave their systematic investigation to future work.

The other logical possibility to consider under the extreme conditions in a NS core relate to QCD deconfinement. In the core of NSs, hadronic matter may undergo a phase transition and deconfine, finding at high densities matter with free quarks and gluons, making up the so-called hybrid stars~\cite{Freedman:1977gz,Schertler:2000xq,Alford:2004pf}. Such a possibility requires a detailed treatment of the EOS, which will shift from hadronic to quark matter as baryon density rises within the star.

The presence of new degrees of freedom would, of course, require the inclusion of new cooling processes~\cite{Iwamoto:1982zz} and, more importantly, could lead to quite general limits on the axion, due to its model-independent coupling to gluons. While the \texttt{MUSES CE} does include tools to deal with hybrid stars and we have performed a first approach to this study, the full analysis of these objects in the context of axion physics lies beyond the scope of the current manuscript and we leave to be explored in future work~\cite{Arias-Aragon:future}.

\section{Stellar mass-radius and cooling curves}\label{sec:SeqCool}

Now that we have established the different compositions to be considered in this study, we must check whether EOS including the different set of particles are in agreement with constraints or not. In order to do so, we will follow roughly the same philosophy as in~\cite{Buschmann:2021juv}: we will use the constraints provided in~\cite{Miller:2021qha} and require that our EOSs result in mass-radius curves that enter, at least, within the 90\% confidence region. In contrast to what was done in\cite{Buschmann:2021juv}, we not only consider the band shown in Fig.~11 of~\cite{Miller:2021qha}, which includes only the results for one of the three models considered in that work, but also include the full result spanning all three models at $1.4M_\odot$ and $2.08M_\odot$. We do this for consistency, as one should consider the dependence on the NS core modelling given in~\cite{Miller:2021qha} which would translate into a slightly larger EOS dependence on the derived axion limit.

\begin{figure}[t!]
    \centering
    \includegraphics[width=0.7 \linewidth,trim=0.65 1 0.5 0.75,clip]{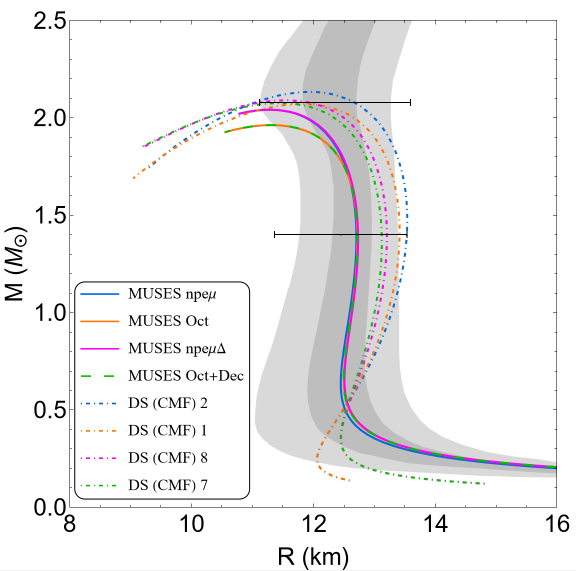}
    \caption{Mass-radius relation for the EOSs obtained with the \texttt{MUSES CE} in solid or dashed and DS~(CMF) in dot-dashed. Blue lines represent equations with \textit{$npe\mu$} composition (\textit{$npe$} for DS~(CMF) 2), orange ones relate to the \textit{Delta} scenario (without muons for DS~(CMF) 8), pink shows the \textit{Octet} composition and, finally, green curves represent the \textit{Octet+Decuplet}. The dark (light) gray bands show the 90\% (50\%) CL mass-radius allowed region by~\cite{Miller:2021qha} using one specific model to describe the NS core. The two gray points, at $1.4M_\odot$ and $2.08M_\odot$ respectively, show the 90\% CL results using all three models considered for the NS core in~\cite{Miller:2021qha}.}
    \label{fig:MvsR}
\end{figure}

We recast both the band and the two points obtained from~\cite{Miller:2021qha} in Fig.~\ref{fig:MvsR}, together with the eight curves obtained from the two different EOSs in each of the four scenarios considered. With solid lines (and one dashed for convenience), one can see that all EOSs produced with the \texttt{MUSES CE}, no matter their composition, comply perfectly with the constraints obtained from the different observables studied in~\cite{Miller:2021qha}. While the DS~(CMF) EOSs stray further from the centre of the band, the inclusion of the two extra points seems to indicate they can still be considered, with DS~(CMF) 2 being a borderline case. Thus, we will include them all in the analysis, acting therefore as a control to our \texttt{MUSES CE}-produced EOS.

In Fig.~\ref{fig:MvsR} one can already observe that the mass-radius curve of NSs with \texttt{MUSES CE} EOS is quite stable under changes in the core composition. Although the inclusion of the baryon octet lowers the maximum value of the NS mass, whenever a NS lighter than roughly $1.7M_\odot$ is considered, no difference in its radius is observed. This behaviour, however, is not observed in the DS~(CMF) family, where slightly different radii are inferred roughly at all stellar masses above $0.5M_\odot$.

The information shown so far determines the stellar structure associated with each EOS, before its thermal evolution is computed with \NSCool. The publicly available version of \texttt{NSCool} includes standard neutrino and photon cooling processes but does not account for axion emission. We therefore extended the code by implementing the nucleonic axion emissivities introduced in Sec.~\ref{sec:QCDAxion}, following the expressions provided in the supplementary material of Ref.~\cite{Buschmann:2021juv}. Within this framework, superfluidity and envelope physics are treated as microphysical inputs to the general relativistic thermal evolution problem. The code allows independent choices for neutron singlet pairing, neutron triplet pairing, and proton singlet pairing.

In our work, we consider three superfluidity models as in~\cite{Buschmann:2021juv}: \texttt{000}, corresponding to no superfluidity; \texttt{SFB00}, corresponding to neutron singlet-state paring; and \texttt{SFB0T73}, including both neutron and proton singlet-state pairing. All these three scenarios are already described within the available version of~\NSCool, including both the relevant suppression factors and PBF processes. $\Delta$ resonances are treated as unpaired in all simulations presented here, while hyperons are assumed to have the same energy gap as neutrons and are treated together with them. Indeed, hyperonic pairing remains strongly model dependent and may substantially suppress hyperonic neutrino emission
channels \cite{Takatsuka:2005bp,Vidana:2004rd,Raduta:2017wpp}. Our approximation, however, is reasonable in the present setup, as we find the hyperonic impact on NS cooling to be marginal; nevertheless, further suppression due to superfluidity could slightly shift the detailed cooling curves.

The outer envelope is treated as a thin external layer rather than explicitly evolved.  In this work, we described the outer layers of the NS using the accreted-envelope model based on Potekhin--Chabrier--Yakovlev relations \cite{Potekhin:1997mn}, already implemented in \NSCool. In this prescription, the envelope contains a light element layer of mass $\Delta M$ above heavier material, and its thickness is parametrized by 
\begin{equation*}
    \eta \equiv g_{s14}^{\,2}\frac{\Delta M}{M} = \frac{P_{\rm light}} {1.193\times 10^{34}\ {\rm dyn\cdot cm^{-2}}}.
\end{equation*}
Here, $g_{s14}=g_s/(10^{14}\,{\rm cm\cdot s^{-2}})$ with $g_s=2.43\times10^{14}\cm\cdot s^{-2}
$ the NS surface gravity, while $P_{\rm light}$ represents the pressure at the bottom of the light element layer. The parameter $\Delta M$, therefore, controls the thermal transparency of the envelopes. This means that larger values correspond to a thicker light element layer, leading to a higher effective surface temperature.

\begin{figure}[t!]
    \centering
    \includegraphics[width=0.68 \linewidth,trim=1.75 1.25 1.4 1,clip]{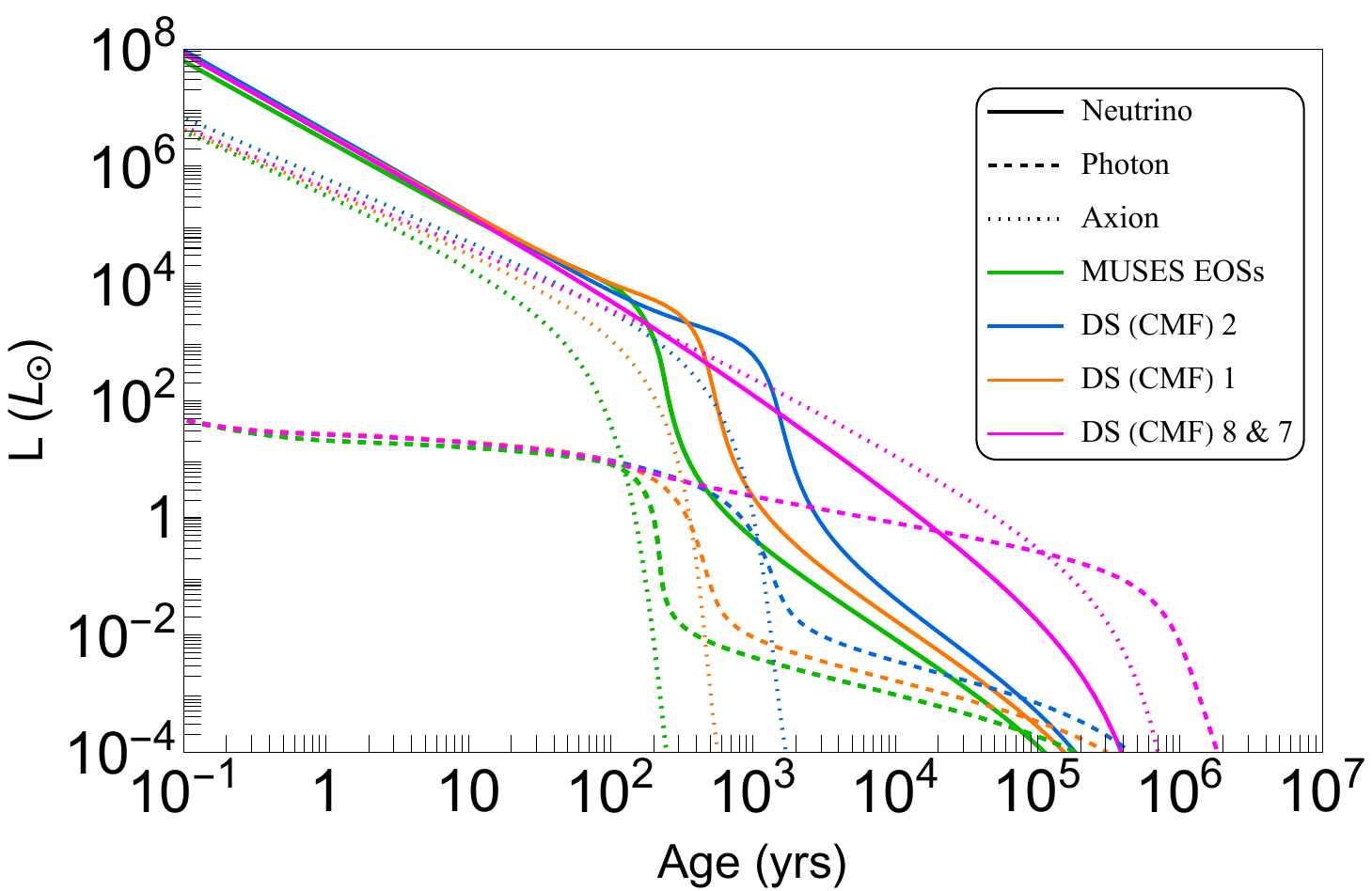}
    \caption{Neutrino (solid), photon (dashed), and axion (dotted) luminosity curves for a star of mass $1.4M_\odot$, envelope characterized by $\Delta M/M=10^{-13}$ and no superfluidity. The axion luminosity was computed for a KSVZ axion with the best-fit mass $m_a=13.61\,\mathrm{meV}$, corresponding to the minimum value of $\chi^2$. Green lines represent all EOSs calculated with the \texttt{MUSES CE}, which coincide among each other, while blue, orange and pink lines correspond to DS~(CMF) 2, 1 and 7 and 8, with the last two also superimposed.}
    \label{fig:Lumi}
\end{figure}

In Fig.~\ref{fig:Lumi} we show the photon, neutrino and axion luminosities for each of the eight EOSs considered, obtained with our full code. We do this for the case of a star with $M=1.4M_\odot$, an envelope characterized by $\Delta M/M=10^{-13}$ and no superfluidity, for a KSVZ axion of best-fit mass $m_a=13.61$ meV. For the representative hyperonic cooling models, we also verified that the combined contribution of the implemented hyperonic modified Urca and bremsstrahlung processes remains subdominant with respect to the total neutrino luminosity over the age range relevant to the analysis. This result shows that, despite the mass-radius curves obtained with different particle contents may be quite similar as depicted in Fig.~\ref{fig:MvsR}, the thermal evolution of each star can indeed be quite different. As a matter of fact, it is easy to see in Fig.~\ref{fig:Lumi} that initial luminosities are roughly the same for all cases, but as the star grows older, certain processes, in particular the fast DURCA emission, turn on at different times depending on the star composition. It is also interesting that the relevant NS age of $\sim10^5$ years identified in~\cite{Buschmann:2021juv} where axion luminosity is maximally relevant also depends largely on the core composition, as allowing for certain particles to be present may reduce the relative axion luminosity and thus require for NSs of different age in order to optimize the limit.

One must, however, be careful when interpreting this result. These cooling curves depend largely on parameters like the NS mass, its abundance of light elements and the superfluidity model considered for its core, not only on the EOS and its particle content. This is why, in the next section, in order to obtain the limits on the axion couplings, we perform a statistical analysis that marginalizes over all nuisance parameters.

\section{Results}\label{sec:Results}

In this section, we present the results we find after revisiting the limits from~\cite{Buschmann:2021juv} on the axion mass in the KSVZ and DFSZ models using the different scenarios introduced for NS core particle content. We used the same data of ages and luminosities of NS as in~\cite{Buschmann:2021juv}, as shown in Table~\ref{table:Stars}.

We then performed a statistical analysis following the strategy introduced in~\cite{Buschmann:2021juv}. For each microscopic input model, we compare the observed ages and luminosities of the selected sources with dedicated cooling simulations and construct a likelihood test statistic from the difference between the best-fit axion and no-axion (null) hypotheses, profiling over the stellar mass and $\Delta M$. The Monte Carlo distribution of the test statistic under the null hypothesis is then used to assign the corresponding $p$-value and significance, while the 95\% upper limit on the axion mass is obtained from the profiled likelihood scan.

\begin{table}[h!]
\centering
\begin{tabular}{|c|c|c|c|}
\hline
Name & $L_\gamma^\infty (10^{33}\ \textrm{erg/s})$ & Age ($10^6$ yr) & Refs\\
\hline
J1856 & $0.065\pm 0.015$ & $0.42\pm 0.08 $ & ~\cite{Mignani:2012mm,Ho:2006uk,Sartore:2012fk}\\
J1308 & $0.32\pm 0.06$ & $0.55\pm 0.25$ & ~\cite{Motch:2009nq,Hambaryan:2011bu}\\
J0720 & $0.22\pm0.11$ & $0.85\pm0.15$ & ~\cite{Tetzlaff:2011kh,Hambaryan:2017wvm}\\
J1605 & $0.4\pm0.1$ & $0.44\pm0.07$ & ~\cite{Tetzlaff:2012rz,Pires:2019qsk}\\
J0659 & $0.28\pm0.14$ & $0.35\pm0.044$ &  ~\cite{suzuki2021quantitative,zharikov2021psr}\\ \hline

\end{tabular}
\caption{List of the NSs considered, together with their photon luminosity $L_\gamma^\infty$ and their age.}\label{table:Stars}
\end{table}\vspace{-0.18cm}

Additionally, we use an adaptive mass grid to refine the region relevant for the best fit, the upper limit, and the likelihood prescription. In particular, we used a logarithmic likelihood, comparing the cooling curves to the data in $\log \mathcal{L}$, in contrast to the linear one used in the previous work. This choice is justified for the analysis of neutron star cooling data, since luminosities and ages are positive quantities spanning several orders of magnitude, and it greatly improves numerical stability.

For the KSVZ axion model, the analysis is performed as a one-dimensional scan in the axion mass $m_a$: the axion-nucleon couplings $g_{aNN}$ from Eq.~\eqref{eq:gaNN} are fixed by Eq.~\eqref{eq:KSVZ}, once $m_a$ is specified and Eq.~\eqref{eq:ma} is used. We perform this analysis for each choice of superfluidity model, EOS family and NS core particle content, and present in Table~\ref{tab:results_KSVZ} the 95\% CL upper limit set on the axion mass, the best-fit value for $m_a$ and its significance $\sigma$, and the $\chi^2$ for the null hypothesis.

In the results shown in Table~\ref{tab:results_KSVZ} we can see that there is no strong dependence on the upper limit on the axion mass due to the particle content allowed to be present in the NS core, which reinforces the robustness of the limit offered by~\cite{Buschmann:2021juv}. While some slight variation is observed in the results of the DS~(CMF) family of EOSs, the dependence on composition is well below that of, for example, the dependence on superfluidity assumptions. All the results we find are consistent in order of magnitude with those shown in the original work~\cite{Buschmann:2021juv}.

One can also observe in our Table~\ref{tab:results_KSVZ} a clear difference from those from the previous analysis: while in Ref.~\cite{Buschmann:2021juv} all best-fit values for the axion mass were negative, here we find both negative and positive values, and overall much closer to zero. The reason for the appearance of negative values, as explained in the original work, is that negative values for the axion mass are allowed to account for mismodelling of the NS, in such a way that the axion emissivity for such mass was negative. The fact that all best-fit values in the original work were negative seems to point in the direction of neutrino cooling being too large. In this work, we incorporate the corrections to the MURCA and bremsstrahlung processes from~\cite{Bottaro:2024ugp}, which effectively reduce neutrino luminosity. After this correction, we find best-fit values for the axion mass which oscillate closely around zero, as one would expect when the null hypothesis already seems to describe observations quite well.

For the DFSZ axion, the same procedure is repeated on a two-parameter grid, since the axion-nucleon couplings $g_{aNN}$ depend not only on $m_a$ but also on $\tan \beta$ as described in Eq.~\eqref{eq:DFSZ}. We then perform, as for the KSVZ model, the analysis independently for each EOS family, core composition and superfluidity, in order to obtain $m_a^{95}$, the 95\% CL upper limit on the axion mass.

\begin{figure*}[t]
    \centering
    \includegraphics[width=0.7 \textwidth,trim=0.5 0.75 0.5 1,clip]{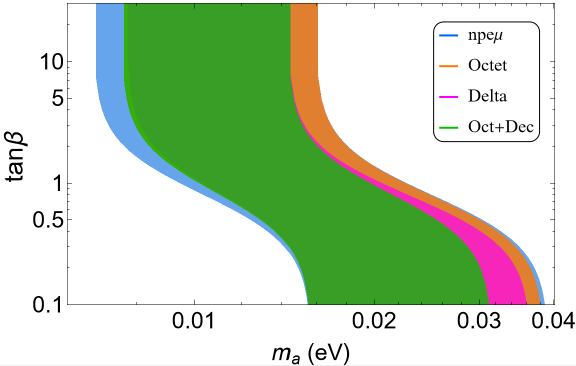}
    \caption{95\% CL upper limit on the axion mass, $m_a^{95}$, shown as shaded regions. Each colour shows the extension where each core composition can set the upper limit, with blue, orange, pink and green representing the $npe\mu$, \textit{Octet}, \textit{Delta} and \textit{Oct+Dec} scenarios.}
    \label{fig:DFSZCompo}
\end{figure*}

The results obtained for DFSZ axions in Fig.~\ref{fig:DFSZCompo} show no strong dependence on the particles present within the NS core. All bands are around the same axion masses, their width given by the variation of superfluidity model and EOS family.

\begin{figure*}[t]
    \centering
    \includegraphics[width=0.7 \textwidth,trim=0.5 1.25 0.5 1,clip]{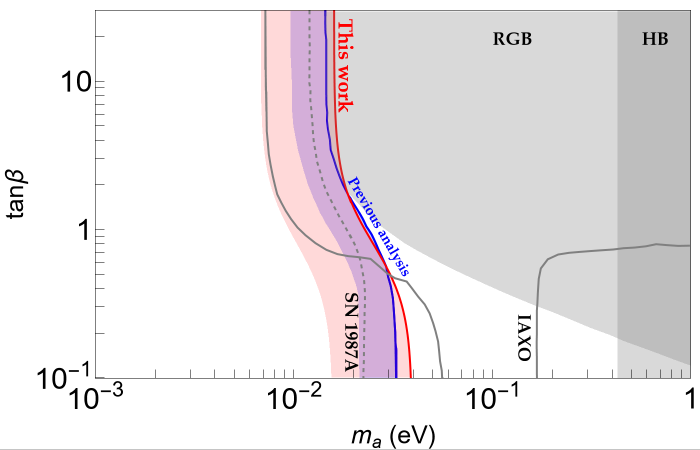}
    \caption{New limit on the mass of DFSZ axions (red) at 95\% CL including all the scenarios and EOSs considered. For comparison, in blue we show the previous analysis from~\cite{Buschmann:2021juv}, and for completeness we include also the limits from red giant branch (RGB)~\cite{Viaux:2013lha,Straniero:2018fbv} and horizontal branch (HB)~\cite{Ayala:2014pea} as shaded gray regions, the SN 1987A~\cite{Carenza:2019pxu} as a dashed gray line. The prospected reach for IAXO~\cite{IAXO:2019mpb} is also shown with a solid gray line.}
    \label{fig:DFSZ_Final}
\end{figure*}

Finally, in Fig.~\ref{fig:DFSZ_Final} we show our full result for the DFSZ axion, including all compositions, both EOS families and all superfluidity models, as a red band. The weakest limit, given by the DS~(CMF) 2 EOS with no superfluidity, is shown as a solid red line.

Overall, our result is compatible with that from~\cite{Buschmann:2021juv}, with our weakest bound being just very slightly looser than the original one, which once again gives further soundness to the bound. On the other hand, our strongest result, given by the \texttt{MUSES CE} EOS with neutron and proton singlet superfluidity in the \textit{$npe\mu$} case, is almost a factor of 2 stronger than that in the previous work. Interestingly enough, it goes right beyond IAXO sensitivity~\cite{IAXO:2019mpb}, covering the window commented in the previous study. If IAXO were to discover an axion in this region, it may then help disentangle the composition within the NS, as this region is precisely sensitive to the particles within the core.

\begin{table*}[t]
\centering\large
\begin{tabular}{|c|ccc|}
\hline
Model 
& $0$-$0$-$0$ 
& SFB-$0$-$0$ 
& SFB-$0$-T73
\\
\hline

$npe\mu$
&
\makecell[c]{
$m^{95}_a= 13.46\,\mathrm{meV}$\\
$\widehat m_a=4.13\,\mathrm{meV}$\\
$\sigma=0.85$\\
$\chi^2=1.16$
}
&
\makecell[c]{
$m^{95}_a=9.71\,\mathrm{meV}$\\
$\widehat m_a=0.22\,\mathrm{meV}$\\
$\sigma=0.01$\\
$\chi^2=0.57$
}
&
\makecell[c]{
$m^{95}_a=7.64\,\mathrm{meV}$\\
$\widehat m_a=-2.54\,\mathrm{meV}$\\
$\sigma=0.22$\\
$\chi^2=0.81$
}
\\
\hline

\textit{Octet}
&
\makecell[c]{
$m^{95}_a=13.46\,\mathrm{meV}$\\
$\widehat m_a=4.12\,\mathrm{meV}$\\
$\sigma=0.85$\\
$\chi^2=1.17$
}
&
\makecell[c]{
$m^{95}_a=9.80\,\mathrm{meV}$\\
$\widehat m_a=0.38\,\mathrm{meV}$\\
$\sigma=0.01$\\
$\chi^2=0.58$
}
&
\makecell[c]{
$m^{95}_a=7.65\,\mathrm{meV}$\\
$\widehat m_a=-2.49\,\mathrm{meV}$\\
$\sigma=0.13$\\
$\chi^2=0.81$
}
\\
\hline

\textit{Delta}
&
\makecell[c]{
$m^{95}_a=13.49\,\mathrm{meV}$\\
$\widehat m_a=4.18\,\mathrm{meV}$\\
$\sigma=0.83$\\
$\chi^2=1.15$
}
&
\makecell[c]{
$m^{95}_a=9.80\,\mathrm{meV}$\\
$\widehat m_a=-0.77\,\mathrm{meV}$\\
$\sigma=0.04$\\
$\chi^2=0.58$
}
&
\makecell[c]{
$m^{95}_a=7.70\,\mathrm{meV}$\\
$\widehat m_a=-2.55\,\mathrm{meV}$\\
$\sigma=0.21$\\
$\chi^2=0.83$
}
\\
\hline

\textit{Oct+Dec}
&
\makecell[c]{
$m^{95}_a=13.46\,\mathrm{meV}$\\
$\widehat m_a=4.12\,\mathrm{meV}$\\
$\sigma=0.85$\\
$\chi^2=1.17$
}
&
\makecell[c]{
$m^{95}_a=9.80\,\mathrm{meV}$\\
$\widehat m_a=0.41\,\mathrm{meV}$\\
$\sigma=0.02$\\
$\chi^2=0.58$
}
&
\makecell[c]{
$m^{95}_a=7.65\,\mathrm{meV}$\\
$\widehat m_a=-2.36\,\mathrm{meV}$\\
$\sigma=0.13$\\
$\chi^2=0.81$
}
\\
\hline

DS~(CMF) 2
&
\makecell[c]{
$m^{95}_a=18.10\,\mathrm{meV}$\\
$\widehat m_a=6.07\,\mathrm{meV}$\\
$\sigma=0.36$\\
$\chi^2=0.50$
}
&
\makecell[c]{
$m^{95}_a=14.83\,\mathrm{meV}$\\
$\widehat m_a=4.21\,\mathrm{meV}$\\
$\sigma=0.87$\\
$\chi^2=1.07$
}
&
\makecell[c]{
$m^{95}_a=13.59\,\mathrm{meV}$\\
$\widehat m_a=4.24\,\mathrm{meV}$\\
$\sigma=0.41$\\
$\chi^2=0.96$
}
\\
\hline
DS~(CMF) 1
&
\makecell[c]{
$m^{95}_a=17.59\,\mathrm{meV}$\\
$\widehat m_a=5.37\,\mathrm{meV}$\\
$\sigma=0.49$\\
$\chi^2=0.60$
}
&
\makecell[c]{
$m^{95}_a=14.74\,\mathrm{meV}$\\
$\widehat m_a=3.34\,\mathrm{meV}$\\
$\sigma=0.45$\\
$\chi^2=0.50$
}
&
\makecell[c]{
$m^{95}_a=12.83\,\mathrm{meV}$\\
$\widehat m_a=2.92\,\mathrm{meV}$\\
$\sigma=0.39$\\
$\chi^2=0.73$
}
\\
\hline

DS~(CMF) 8
&
\makecell[c]{
$m^{95}_a=16.68\,\mathrm{meV}$\\
$\widehat m_a=5.80\,\mathrm{meV}$\\
$\sigma=0.66$\\
$\chi^2=0.46$
}
&
\makecell[c]{
$m^{95}_a=13.62\,\mathrm{meV}$\\
$\widehat m_a=3.29\,\mathrm{meV}$\\
$\sigma=0.73$\\
$\chi^2=0.60$
}
&
\makecell[c]{
$m^{95}_a=12.09\,\mathrm{meV}$\\
$\widehat m_a=3.30\,\mathrm{meV}$\\
$\sigma=0.29$\\
$\chi^2=0.69$
}
\\

\hline
DS~(CMF) 7
&
\makecell[c]{
$m^{95}_a=16.27\,\mathrm{meV}$\\
$\widehat m_a=5.13\,\mathrm{meV}$\\
$\sigma=0.84$\\
$\chi^2=0.57$
}
&
\makecell[c]{
$m^{95}_a=13.61\,\mathrm{meV}$\\
$\widehat m_a=2.23\,\mathrm{meV}$\\
$\sigma=0.25$\\
$\chi^2=0.17$
}
&
\makecell[c]{
$m^{95}_a=12.68\,\mathrm{meV}$\\
$\widehat m_a=2.65\,\mathrm{meV}$\\
$\sigma=0.20$\\
$\chi^2=0.41$
}
\\
\hline
\end{tabular}
\caption{Results for the KSVZ axion case, showing the 95\% CL upper limit on the axion mass, $m_a^{95}$, the best-fit axion mass $\hat{m}_a$ and its significance $\sigma$, and the null model $\chi^2$ value.}
\label{tab:results_KSVZ}
\end{table*}

\clearpage

\section{Conclusions} \label{sec:Conclusions}

In this work, we performed a new analysis of the QCD axion mass limit obtained from NS cooling by varying the particle content within the core of the star. We defined four scenarios according to which particles may appear in the NS core: $npe\mu$, \textit{Octet}, \textit{Delta}, and \textit{Oct+Dec}. We then obtained the EOSs using the \texttt{MUSES CE}~\cite{MusesCalculationEngine} for each of these four cases, and as a control also used four EOSs found within the repository CompOSE~\cite{CompOSECoreTeam:2022ddl,CompOSEweb} with analogous compositions. We largely upgraded the cooling code~\NSCool $\ $with the inclusion of all relevant new processes for neutrino emission depending on the particle content considered.

After checking in Fig.~\ref{fig:MvsR} that all these equations of state comply with current limits on NS mass and radii, we used the open source code~\NSCool $\ $to time-evolve the NSs and obtain their thermal history as shown in Fig.~\ref{fig:Lumi}. We then performed the same statistical analysis as in~\cite{Buschmann:2021juv} in order to study dependence on the NS core composition under the same conditions. We did so for the two main QCD axion benchmark scenarios, namely the KSVZ and DFSZ invisible axion models.

In the KSVZ axion framework we find that our results, presented in Table~\ref{tab:results_KSVZ}, are perfectly compatible with those offered in the first study~\cite{Buschmann:2021juv}, adding to the stability of this limit. Allowing for hyperons and/or spin $\frac{3}{2}$ resonances in the NS core does not seem to largely affect the limit derived on the axion mass, with EOS family and superfluidity having a stronger impact. Thanks to the correction to neutrino bremsstrahlung and MURCA processes from~\cite{Bottaro:2024ugp}, we find a best-fit value for the axion mass that oscillates around zero, with both positive and negative values. This contrasts with the always negative and rather large values found in the original work of~\cite{Buschmann:2021juv}.

In the context of DFSZ axions, we explore the plane spanned by the axion mass $m_a$ and the extra parameter $\tan\beta$ and show the composition dependence in Fig.~\ref{fig:DFSZCompo}. There, the same conclusion as for the KSVZ scenario is drawn: allowing the presence of heavier SM particles in the NS core has a mild effect on the derived constraint on the axion mass. In Fig.~\ref{fig:DFSZ_Final} we offer a clear comparison with the result of the first study and ours. Our new analysis shows again clear agreement with the previous one, without worsening the most pessimistic limit, which once again strengthens the claim of this limit. On the contrary, the strongest limit is enhanced by almost a factor of two, reaching now beyond the expected sensitivity of IAXO. This implies that an axion discovery in this window would allow us to test which particles may or may not lie in the heart of neutron stars.

In addition to the results presented here, this work shows the versatility of the \texttt{MUSES CE}~\cite{MusesCalculationEngine} and opens the door to exciting future analyses, both in SM and BSM scenarios. Of particular interest for us is the case of hybrid stars, with a quark-gluon phase present in its core. Although we have already started a preliminary study, it requires a different framework with regard to neutrino and axion emission, and therefore leave it to be explored in the future~\cite{Arias-Aragon:future}.

\section*{Acknowledgements} \label{sec:Ack}

The authors of this work are deeply grateful to M. Giannotti, L. Merlo and E. Nardi for encouraging the development of this project from its early stages as well as for enlightening conversations. The authors are also thankful for their hospitality during different moments in the past year, for welcoming us to their institutions, and for allowing us to discuss this matter with a broader audience about. We kindly thank as well the MUSES Collaboration, for their openness, availability, and swiftness, be it answering doubts or checking bugs in the code.

\footnotesize

\bibliography{biblio}{}

\providecommand{\href}[2]{#2}\begingroup\raggedright\begin{thebibliography}{100}

\bibitem{Lattimer:2004sa}
J.~M. Lattimer and M.~Prakash, {\it {The Ultimate energy density of observable cold matter}},  Phys. Rev. Lett. {\bf 94} (2005) 111101, [\href{http://arxiv.org/abs/astro-ph/0411280}{{\tt astro-ph/0411280}}].

\bibitem{Page:1997mj}
D.~Page, {\it {Thermal evolution of isolated neutron stars}},  in {\em {NATO Advanced Study Institute: The Many Faces of Neutron Stars}}, 6, 1997.
\newblock \href{http://arxiv.org/abs/astro-ph/9706259}{{\tt astro-ph/9706259}}.

\bibitem{Yakovlev:2000jp}
D.~G. Yakovlev, A.~D. Kaminker, O.~Y. Gnedin, and P.~Haensel, {\it {Neutrino emission from neutron stars}},  Phys. Rept. {\bf 354} (2001) 1, [\href{http://arxiv.org/abs/astro-ph/0012122}{{\tt astro-ph/0012122}}].

\bibitem{Page:2005fq}
D.~Page, U.~Geppert, and F.~Weber, {\it {The Cooling of compact stars}},  Nucl. Phys. A {\bf 777} (2006) 497--530, [\href{http://arxiv.org/abs/astro-ph/0508056}{{\tt astro-ph/0508056}}].

\bibitem{Maxwell:1986pj}
O.~V. Maxwell, {\it {Neutrino Emission Processes in Hyperon Populated Neutron Stars}},  Astrophys. J. {\bf 316} (1987) 691--707.

\bibitem{Prakash:1992zng}
M.~Prakash, M.~Prakash, J.~M. Lattimer, and C.~J. Pethick, {\it {Rapid cooling of neutron stars by hyperons and Delta isobars}},  Astrophys. J. Lett. {\bf 390} (1992) L77.

\bibitem{Muto:1994unh}
T.~Muto, T.~Tatsumi, and N.~Iwamoto, {\it {Axion emission from meson condensates}},  Phys. Rev. D {\bf 50} (1994) 6089--6099.

\bibitem{Ivanenko:1969bs}
D.~D. Ivanenko and D.~F. Kurdgelaidze, {\it {Quark stars}},  Sov. Phys. J. {\bf 13} (1970) 1015--1019.

\bibitem{Iwamoto:1982zz}
N.~Iwamoto, {\it {Neutrino emissivities and mean free paths of degenerate quark matter}},  Annals Phys. {\bf 141} (1982) 1--49.

\bibitem{Lattimer:2004pg}
J.~M. Lattimer and M.~Prakash, {\it {The physics of neutron stars}},  Science {\bf 304} (2004) 536--542, [\href{http://arxiv.org/abs/astro-ph/0405262}{{\tt astro-ph/0405262}}].

\bibitem{Yakovlev_2004}
D.~Yakovlev and C.~Pethick, {\it Neutron star cooling},  Annual Review of Astronomy and Astrophysics {\bf 42} (Sept., 2004) 169–210.

\bibitem{Cruz-Camacho:2024odu}
{\bf MUSES} Collaboration, N.~Cruz-Camacho, R.~Kumar, M.~Reinke~Pelicer, J.~Peterson, T.~A. Manning, R.~Haas, V.~Dexheimer, and J.~Noronha-Hostler, {\it {Phase stability in the three-dimensional open-source code for the chiral mean-field model}},  Phys. Rev. D {\bf 111} (2025), no.~9 094030, [\href{http://arxiv.org/abs/2409.06837}{{\tt arXiv:2409.06837}}].

\bibitem{Peccei:1977hh}
R.~D. Peccei and H.~R. Quinn, {\it {CP Conservation in the Presence of Instantons}},  Phys. Rev. Lett. {\bf 38} (1977) 1440--1443.

\bibitem{Peccei:1977ur}
R.~D. Peccei and H.~R. Quinn, {\it {Constraints Imposed by CP Conservation in the Presence of Instantons}},  Phys. Rev. D {\bf 16} (1977) 1791--1797.

\bibitem{Wilczek:1977pj}
F.~Wilczek, {\it {Problem of Strong $P$ and $T$ Invariance in the Presence of Instantons}},  Phys. Rev. Lett. {\bf 40} (1978) 279--282.

\bibitem{Weinberg:1977ma}
S.~Weinberg, {\it {A New Light Boson?}},  Phys. Rev. Lett. {\bf 40} (1978) 223--226.

\bibitem{Fiorillo:2025zzx}
D.~F.~G. Fiorillo, A.~Lella, C.~A.~J. O'Hare, and E.~Vitagliano, {\it {Leading Bounds on Micrometer to Picometer Fifth Forces from Neutron Star Cooling}},  Phys. Rev. Lett. {\bf 135} (2025), no.~21 211003, [\href{http://arxiv.org/abs/2506.19906}{{\tt arXiv:2506.19906}}].

\bibitem{Fiorillo:2026wso}
D.~F.~G. Fiorillo, A.~Lella, G.~G. Raffelt, N.~Selimovic, and E.~Vitagliano, {\it {Neutron Star Bounds on Muonic Fifth Forces from Picometer to Kilometer Scales}},  \href{http://arxiv.org/abs/2605.24094}{{\tt arXiv:2605.24094}}.

\bibitem{Bramante:2023djs}
J.~Bramante and N.~Raj, {\it {Dark matter in compact stars}},  Phys. Rept. {\bf 1052} (2024) 1--48, [\href{http://arxiv.org/abs/2307.14435}{{\tt arXiv:2307.14435}}].

\bibitem{Dey:2025atz}
U.~K. Dey and S.~Gope, {\it {Asymmetric cannibal dark matter: constraints from neutron star}},  JHEP {\bf 04} (2026) 098, [\href{http://arxiv.org/abs/2509.06892}{{\tt arXiv:2509.06892}}].

\bibitem{Hao:2025fea}
Y.~Hao and Z.~Chen, {\it {Dark photons from dineutron decays in neutron stars}},  Phys. Rev. D {\bf 112} (2025), no.~7 075017, [\href{http://arxiv.org/abs/2504.05447}{{\tt arXiv:2504.05447}}].

\bibitem{Abel:2020pzs}
C.~Abel {\em et.~al.}, {\it {Measurement of the Permanent Electric Dipole Moment of the Neutron}},  Phys. Rev. Lett. {\bf 124} (2020), no.~8 081803, [\href{http://arxiv.org/abs/2001.11966}{{\tt arXiv:2001.11966}}].

\bibitem{Preskill:1982cy}
J.~Preskill, M.~B. Wise, and F.~Wilczek, {\it {Cosmology of the Invisible Axion}},  Phys. Lett. B {\bf 120} (1983) 127--132.

\bibitem{Abbott:1982af}
L.~F. Abbott and P.~Sikivie, {\it {A Cosmological Bound on the Invisible Axion}},  Phys. Lett. B {\bf 120} (1983) 133--136.

\bibitem{Dine:1982ah}
M.~Dine and W.~Fischler, {\it {The Not So Harmless Axion}},  Phys. Lett. B {\bf 120} (1983) 137--141.

\bibitem{Carenza:2019pxu}
P.~Carenza, T.~Fischer, M.~Giannotti, G.~Guo, G.~Mart{\'\i}nez-Pinedo, and A.~Mirizzi, {\it {Improved axion emissivity from a supernova via nucleon-nucleon bremsstrahlung}},  JCAP {\bf 10} (2019), no.~10 016, [\href{http://arxiv.org/abs/1906.11844}{{\tt arXiv:1906.11844}}]. [Erratum: JCAP 05, E01 (2020)].

\bibitem{Viaux:2013lha}
N.~Viaux, M.~Catelan, P.~B. Stetson, G.~Raffelt, J.~Redondo, A.~A.~R. Valcarce, and A.~Weiss, {\it {Neutrino and axion bounds from the globular cluster M5 (NGC 5904)}},  Phys. Rev. Lett. {\bf 111} (2013) 231301, [\href{http://arxiv.org/abs/1311.1669}{{\tt arXiv:1311.1669}}].

\bibitem{Straniero:2018fbv}
O.~Straniero, I.~Dominguez, M.~Giannotti, and A.~Mirizzi, {\it {Axion-electron coupling from the RGB tip of Globular Clusters}},  in {\em {13th Patras Workshop on Axions, WIMPs and WISPs}}, pp.~172--176, 2018.
\newblock \href{http://arxiv.org/abs/1802.10357}{{\tt arXiv:1802.10357}}.

\bibitem{Leinson:2014ioa}
L.~B. Leinson, {\it {Axion mass limit from observations of the neutron star in Cassiopeia A}},  JCAP {\bf 08} (2014) 031, [\href{http://arxiv.org/abs/1405.6873}{{\tt arXiv:1405.6873}}].

\bibitem{Hamaguchi:2018oqw}
K.~Hamaguchi, N.~Nagata, K.~Yanagi, and J.~Zheng, {\it {Limit on the Axion Decay Constant from the Cooling Neutron Star in Cassiopeia A}},  Phys. Rev. D {\bf 98} (2018), no.~10 103015, [\href{http://arxiv.org/abs/1806.07151}{{\tt arXiv:1806.07151}}].

\bibitem{Sedrakian:2015krq}
A.~Sedrakian, {\it {Axion cooling of neutron stars}},  Phys. Rev. D {\bf 93} (2016), no.~6 065044, [\href{http://arxiv.org/abs/1512.07828}{{\tt arXiv:1512.07828}}].

\bibitem{Carenza:2024ehj}
P.~Carenza, M.~Giannotti, J.~Isern, A.~Mirizzi, and O.~Straniero, {\it {Axion astrophysics}},  Phys. Rept. {\bf 1117} (2025) 1--102, [\href{http://arxiv.org/abs/2411.02492}{{\tt arXiv:2411.02492}}].

\bibitem{Keller:2012yr}
J.~Keller and A.~Sedrakian, {\it {Axions from cooling compact stars}},  Nucl. Phys. A {\bf 897} (2013) 62--69, [\href{http://arxiv.org/abs/1205.6940}{{\tt arXiv:1205.6940}}].

\bibitem{Sedrakian:2018kdm}
A.~Sedrakian, {\it {Axion cooling of neutron stars. II. Beyond hadronic axions}},  Phys. Rev. D {\bf 99} (2019), no.~4 043011, [\href{http://arxiv.org/abs/1810.00190}{{\tt arXiv:1810.00190}}].

\bibitem{Yadav:2025tmq}
S.~Yadav, M.~Mishra, and T.~G. Sarkar, {\it {Magnetic field and EOS effects on Axion emission from the non-rotating Neutron Stars}},  J. Subatomic Part. Cosmol. {\bf 4} (2025) 100175.

\bibitem{Gomez-Banon:2024oux}
A.~G{\'o}mez-Ba{\~n}{\'o}n, K.~Bartnick, K.~Springmann, and J.~A. Pons, {\it {Constraining Light QCD Axions with Isolated Neutron Star Cooling}},  Phys. Rev. Lett. {\bf 133} (2024), no.~25 251002, [\href{http://arxiv.org/abs/2408.07740}{{\tt arXiv:2408.07740}}].

\bibitem{Fiorillo:2025gnd}
D.~F.~G. Fiorillo, {\'A}.~Gil~Muyor, H.-T. Janka, G.~G. Raffelt, and E.~Vitagliano, {\it {Axion-photon conversion in transient compact stars: Systematics, constraints, and opportunities}},  JCAP {\bf 03} (2026) 053, [\href{http://arxiv.org/abs/2509.13322}{{\tt arXiv:2509.13322}}].

\bibitem{Fiorillo:2026dqu}
D.~F.~G. Fiorillo, A.~Lella, G.~G. Raffelt, N.~Selimovic, and E.~Vitagliano, {\it {Production of Leptophilic Bosons in Ultradegenerate Relativistic Matter}},  \href{http://arxiv.org/abs/2605.24081}{{\tt arXiv:2605.24081}}.

\bibitem{Kachelriess:1997kn}
M.~Kachelriess, C.~Wilke, and G.~Wunner, {\it {Axion cyclotron emissivity of magnetized white dwarfs and neutron stars}},  Phys. Rev. D {\bf 56} (1997) 1313--1319, [\href{http://arxiv.org/abs/astro-ph/9701056}{{\tt astro-ph/9701056}}].

\bibitem{Leinson:2019cqv}
L.~B. Leinson, {\it {Constraints on axions from neutron star in HESS J1731-347}},  JCAP {\bf 11} (2019) 031, [\href{http://arxiv.org/abs/1909.03941}{{\tt arXiv:1909.03941}}].

\bibitem{Beznogov:2018fda}
M.~V. Beznogov, E.~Rrapaj, D.~Page, and S.~Reddy, {\it {Constraints on Axion-like Particles and Nucleon Pairing in Dense Matter from the Hot Neutron Star in HESS J1731-347}},  Phys. Rev. C {\bf 98} (2018), no.~3 035802, [\href{http://arxiv.org/abs/1806.07991}{{\tt arXiv:1806.07991}}].

\bibitem{Iwamoto:1984ir}
N.~Iwamoto, {\it {Axion Emission from Neutron Stars}},  Phys. Rev. Lett. {\bf 53} (1984) 1198--1201.

\bibitem{Iwamoto:1992jp}
N.~Iwamoto, {\it {Nucleon-nucleon bremsstrahlung of axions and pseudoscalar particles from neutron star matter}},  Phys. Rev. D {\bf 64} (2001) 043002.

\bibitem{Yakovlev:1998wr}
D.~G. Yakovlev, A.~D. Kaminker, and K.~P. Levenfish, {\it {Neutrino emission due to Cooper pairing of nucleons in cooling neutron stars}},  Astron. Astrophys. {\bf 343} (1999) 650, [\href{http://arxiv.org/abs/astro-ph/9812366}{{\tt astro-ph/9812366}}].

\bibitem{ParticleDataGroup:2024cfk}
{\bf Particle Data Group} Collaboration, S.~Navas {\em et.~al.}, {\it {Review of particle physics}},  Phys. Rev. D {\bf 110} (2024), no.~3 030001.

\bibitem{Buschmann:2021juv}
M.~Buschmann, C.~Dessert, J.~W. Foster, A.~J. Long, and B.~R. Safdi, {\it {Upper Limit on the QCD Axion Mass from Isolated Neutron Star Cooling}},  Phys. Rev. Lett. {\bf 128} (2022), no.~9 091102, [\href{http://arxiv.org/abs/2111.09892}{{\tt arXiv:2111.09892}}].

\bibitem{Kim:1979if}
J.~E. Kim, {\it {Weak Interaction Singlet and Strong CP Invariance}},  Phys. Rev. Lett. {\bf 43} (1979) 103.

\bibitem{Shifman:1979if}
M.~A. Shifman, A.~I. Vainshtein, and V.~I. Zakharov, {\it {Can Confinement Ensure Natural CP Invariance of Strong Interactions?}},  Nucl. Phys. B {\bf 166} (1980) 493--506.

\bibitem{Dine:1981rt}
M.~Dine, W.~Fischler, and M.~Srednicki, {\it {A Simple Solution to the Strong CP Problem with a Harmless Axion}},  Phys. Lett. B {\bf 104} (1981) 199--202.

\bibitem{Zhitnitsky:1980tq}
A.~R. Zhitnitsky, {\it {On Possible Suppression of the Axion Hadron Interactions. (In Russian)}},  Sov. J. Nucl. Phys. {\bf 31} (1980) 260.

\bibitem{Cavan-Piton:2024ayu}
M.~Cavan-Piton, D.~Guadagnoli, M.~Oertel, H.~Seong, and L.~Vittorio, {\it {Axion Emission from Strange Matter in Core-Collapse SNe}},  Phys. Rev. Lett. {\bf 133} (2024), no.~12 121002, [\href{http://arxiv.org/abs/2401.10979}{{\tt arXiv:2401.10979}}].

\bibitem{Camalich:2020wac}
J.~M. Camalich, J.~Terol-Calvo, L.~Tolos, and R.~Ziegler, {\it {Supernova Constraints on Dark Flavored Sectors}},  Phys. Rev. D {\bf 103} (2021), no.~12 L121301, [\href{http://arxiv.org/abs/2012.11632}{{\tt arXiv:2012.11632}}].

\bibitem{Vonk:2021sit}
T.~Vonk, F.-K. Guo, and U.-G. Mei{\ss}ner, {\it {The axion-baryon coupling in SU(3) heavy baryon chiral perturbation theory}},  JHEP {\bf 08} (2021) 024, [\href{http://arxiv.org/abs/2104.10413}{{\tt arXiv:2104.10413}}].

\bibitem{MusesCalculationEngine}
T.~A. Manning, {\it Muses calculation engine},  2, 2025.

\bibitem{ReinkePelicer:2025vuh}
M.~Reinke~Pelicer {\em et.~al.}, {\it {Building neutron stars with the MUSES calculation engine}},  Phys. Rev. D {\bf 111} (2025), no.~10 103037, [\href{http://arxiv.org/abs/2502.07902}{{\tt arXiv:2502.07902}}].

\bibitem{Dexheimer:2008ax}
V.~Dexheimer and S.~Schramm, {\it {Proto-Neutron and Neutron Stars in a Chiral SU(3) Model}},  Astrophys. J. {\bf 683} (2008) 943--948, [\href{http://arxiv.org/abs/0802.1999}{{\tt arXiv:0802.1999}}].

\bibitem{Dexheimer:2009hi}
V.~A. Dexheimer and S.~Schramm, {\it {A Novel Approach to Model Hybrid Stars}},  Phys. Rev. C {\bf 81} (2010) 045201, [\href{http://arxiv.org/abs/0901.1748}{{\tt arXiv:0901.1748}}].

\bibitem{Dexheimer:2017nse}
V.~Dexheimer, {\it {Tabulated Neutron Star Equations of State Modeled within the Chiral Mean Field Model}},  Publ. Astron. Soc. Austral. {\bf 34} (2017) E006, [\href{http://arxiv.org/abs/1708.08342}{{\tt arXiv:1708.08342}}].

\bibitem{Dexheimer:2020rlp}
V.~Dexheimer, R.~O. Gomes, T.~Kl{\"a}hn, S.~Han, and M.~Salinas, {\it {GW190814 as a massive rapidly rotating neutron star with exotic degrees of freedom}},  Phys. Rev. C {\bf 103} (2021), no.~2 025808, [\href{http://arxiv.org/abs/2007.08493}{{\tt arXiv:2007.08493}}].

\bibitem{NSCool}
D.~Page, {\it Nscool},  2016.

\bibitem{1994AstL...20...43L}
K.~P. {Levenfish} and D.~G. {Yakovlev}, {\it {Suppression of neutrino energy losses in reactions of direct urca processes by superfluidity in neutron star nuclei}},  Astronomy Letters {\bf 20} (Jan., 1994) 43--51.

\bibitem{Miller:2021qha}
M.~C. Miller {\em et.~al.}, {\it {The Radius of PSR J0740+6620 from NICER and XMM-Newton Data}},  Astrophys. J. Lett. {\bf 918} (2021), no.~2 L28, [\href{http://arxiv.org/abs/2105.06979}{{\tt arXiv:2105.06979}}].

\bibitem{GrillidiCortona:2015jxo}
G.~Grilli~di Cortona, E.~Hardy, J.~Pardo~Vega, and G.~Villadoro, {\it {The QCD axion, precisely}},  JHEP {\bf 01} (2016) 034, [\href{http://arxiv.org/abs/1511.02867}{{\tt arXiv:1511.02867}}].

\bibitem{DiLuzio:2020wdo}
L.~Di~Luzio, M.~Giannotti, E.~Nardi, and L.~Visinelli, {\it {The landscape of QCD axion models}},  Phys. Rept. {\bf 870} (2020) 1--117, [\href{http://arxiv.org/abs/2003.01100}{{\tt arXiv:2003.01100}}].

\bibitem{Negele:1971vb}
J.~W. Negele and D.~Vautherin, {\it {Neutron star matter at subnuclear densities}},  Nucl. Phys. A {\bf 207} (1973) 298--320.

\bibitem{Chamel:2008ca}
N.~Chamel and P.~Haensel, {\it {Physics of Neutron Star Crusts}},  Living Rev. Rel. {\bf 11} (2008) 10, [\href{http://arxiv.org/abs/0812.3955}{{\tt arXiv:0812.3955}}].

\bibitem{Du2019}
X.~Du, A.~W. Steiner, and J.~W. Holt, {\it Hot and dense homogeneous nucleonic matter constrained by observations, experiment, and theory},  Phys. Rev. C {\bf 99} (2019) 025803, [\href{http://arxiv.org/abs/1802.09710}{{\tt arXiv:1802.09710}}].

\bibitem{Du2022}
X.~Du, A.~W. Steiner, and J.~W. Holt, {\it Hot and dense matter equation of state probability distributions for astrophysical simulations},  Phys. Rev. C {\bf 105} (2022) 035803, [\href{http://arxiv.org/abs/2107.06697}{{\tt arXiv:2107.06697}}].

\bibitem{Machleidt:2011zz}
R.~Machleidt and D.~R. Entem, {\it {Chiral effective field theory and nuclear forces}},  Phys. Rept. {\bf 503} (2011) 1--75, [\href{http://arxiv.org/abs/1105.2919}{{\tt arXiv:1105.2919}}].

\bibitem{Weinberg1990}
S.~Weinberg, {\it Nuclear forces from chiral lagrangians},  Physics Letters B {\bf 251} (1990), no.~2 288--292.

\bibitem{Wellenhofer:2014hya}
C.~Wellenhofer, J.~W. Holt, N.~Kaiser, and W.~Weise, {\it {Nuclear thermodynamics from chiral low-momentum interactions}},  Phys. Rev. C {\bf 89} (2014), no.~6 064009, [\href{http://arxiv.org/abs/1404.2136}{{\tt arXiv:1404.2136}}].

\bibitem{Coraggio:2012ca}
L.~Coraggio, J.~W. Holt, N.~Itaco, R.~Machleidt, and F.~Sammarruca, {\it {Reduced regulator dependence of neutron-matter predictions with perturbative chiral interactions}},  Phys. Rev. C {\bf 87} (2013), no.~1 014322, [\href{http://arxiv.org/abs/1209.5537}{{\tt arXiv:1209.5537}}].

\bibitem{Papazoglou:1998vr}
P.~Papazoglou, D.~Zschiesche, S.~Schramm, J.~Schaffner-Bielich, H.~Stoecker, and W.~Greiner, {\it {Nuclei in a chiral SU(3) model}},  Phys. Rev. C {\bf 59} (1999) 411--427, [\href{http://arxiv.org/abs/nucl-th/9806087}{{\tt nucl-th/9806087}}].

\bibitem{Aryal:2020ocm}
K.~Aryal, C.~Constantinou, R.~L.~S. Farias, and V.~Dexheimer, {\it {High-Energy Phase Diagrams with Charge and Isospin Axes under Heavy-Ion Collision and Stellar Conditions}},  Phys. Rev. D {\bf 102} (2020), no.~7 076016, [\href{http://arxiv.org/abs/2004.03039}{{\tt arXiv:2004.03039}}].

\bibitem{Takatsuka:2004ch}
T.~Takatsuka, {\it {Hyperon-mixed neutron stars}},  Prog. Theor. Phys. Suppl. {\bf 156} (2004) 84--103.

\bibitem{Lonardoni:2014bwa}
D.~Lonardoni, A.~Lovato, S.~Gandolfi, and F.~Pederiva, {\it {Hyperon Puzzle: Hints from Quantum Monte Carlo Calculations}},  Phys. Rev. Lett. {\bf 114} (2015), no.~9 092301, [\href{http://arxiv.org/abs/1407.4448}{{\tt arXiv:1407.4448}}].

\bibitem{Bombaci:2016xzl}
I.~Bombaci, {\it {The Hyperon Puzzle in Neutron Stars}},  JPS Conf. Proc. {\bf 17} (2017) 101002, [\href{http://arxiv.org/abs/1601.05339}{{\tt arXiv:1601.05339}}].

\bibitem{Vidana:2019zuz}
I.~Vida{\~n}a, {\it {Do hyperons exist in the neutron star interior?}},  AIP Conf. Proc. {\bf 2130} (2019), no.~1 040011.

\bibitem{Weissenborn:2011ut}
S.~Weissenborn, D.~Chatterjee, and J.~Schaffner-Bielich, {\it {Hyperons and massive neutron stars: vector repulsion and SU(3) symmetry}},  Phys. Rev. C {\bf 85} (2012), no.~6 065802, [\href{http://arxiv.org/abs/1112.0234}{{\tt arXiv:1112.0234}}]. [Erratum: Phys.Rev.C 90, 019904 (2014)].

\bibitem{Astashenok:2014pua}
A.~V. Astashenok, S.~Capozziello, and S.~D. Odintsov, {\it {Maximal neutron star mass and the resolution of the hyperon puzzle in modified gravity}},  Phys. Rev. D {\bf 89} (2014), no.~10 103509, [\href{http://arxiv.org/abs/1401.4546}{{\tt arXiv:1401.4546}}].

\bibitem{Katayama:2015dga}
T.~Katayama and K.~Saito, {\it {Hyperons in neutron stars}},  Phys. Lett. B {\bf 747} (2015) 43--47, [\href{http://arxiv.org/abs/1501.05419}{{\tt arXiv:1501.05419}}].

\bibitem{Raduta:2017wpp}
A.~R. Raduta, A.~Sedrakian, and F.~Weber, {\it {Cooling of hypernuclear compact stars}},  Mon. Not. Roy. Astron. Soc. {\bf 475} (2018), no.~4 4347--4356, [\href{http://arxiv.org/abs/1712.00584}{{\tt arXiv:1712.00584}}].

\bibitem{Vidana:2022tlx}
I.~Vida{\~n}a, {\it {Neutron stars and the hyperon puzzle}},  EPJ Web Conf. {\bf 271} (2022) 09001.

\bibitem{Hanauske:1999ga}
M.~Hanauske, D.~Zschiesche, S.~Pal, S.~Schramm, H.~Stoecker, and W.~Greiner, {\it {Neutron star properties in a chiral SU(3) model}},  Astrophys. J. {\bf 537} (2000) 958, [\href{http://arxiv.org/abs/astro-ph/9909052}{{\tt astro-ph/9909052}}].

\bibitem{Masuda:2012ed}
K.~Masuda, T.~Hatsuda, and T.~Takatsuka, {\it {Hadron{\textendash}quark crossover and massive hybrid stars}},  PTEP {\bf 2013} (2013), no.~7 073D01, [\href{http://arxiv.org/abs/1212.6803}{{\tt arXiv:1212.6803}}].

\bibitem{1995A&A...297..717Y}
D.~G. {Yakovlev} and K.~P. {Levenfish}, {\it {Modified URCA process in neutron star cores.}},  Astronomy and Astrophysics {\bf 297} (May, 1995) 717.

\bibitem{Friman:1979ecl}
B.~L. Friman and O.~V. Maxwell, {\it {Neutron Star Neutrino Emissivities}},  Astrophys. J. {\bf 232} (1979) 541--557.

\bibitem{Bottaro:2024ugp}
S.~Bottaro, A.~Caputo, and D.~F.~G. Fiorillo, {\it {Neutrino emission in cold neutron stars: Bremsstrahlung and modified urca rates reexamined}},  JCAP {\bf 11} (2024) 015, [\href{http://arxiv.org/abs/2406.18640}{{\tt arXiv:2406.18640}}].

\bibitem{Akmal:1998cf}
A.~Akmal, V.~R. Pandharipande, and D.~G. Ravenhall, {\it {The Equation of state of nucleon matter and neutron star structure}},  Phys. Rev. C {\bf 58} (1998) 1804--1828, [\href{http://arxiv.org/abs/nucl-th/9804027}{{\tt nucl-th/9804027}}].

\bibitem{Pearson:2018tkr}
J.~M. Pearson, N.~Chamel, A.~Y. Potekhin, A.~F. Fantina, C.~Ducoin, A.~K. Dutta, and S.~Goriely, {\it {Unified equations of state for cold non-accreting neutron stars with Brussels{\textendash}Montreal functionals {\textendash} I. Role of symmetry energy}},  Mon. Not. Roy. Astron. Soc. {\bf 481} (2018), no.~3 2994--3026, [\href{http://arxiv.org/abs/1903.04981}{{\tt arXiv:1903.04981}}]. [Erratum: Mon.Not.Roy.Astron.Soc. 486, 768 (2019)].

\bibitem{CompOSEweb}
\url{https://compose.obspm.fr/}.

\bibitem{Oertel:2016bki}
M.~Oertel, M.~Hempel, T.~Kl{\"a}hn, and S.~Typel, {\it {Equations of state for supernovae and compact stars}},  Rev. Mod. Phys. {\bf 89} (2017), no.~1 015007, [\href{http://arxiv.org/abs/1610.03361}{{\tt arXiv:1610.03361}}].

\bibitem{CompOSECoreTeam:2022ddl}
{\bf CompOSE Core Team} Collaboration, S.~Typel {\em et.~al.}, {\it {CompOSE Reference Manual}},  Eur. Phys. J. A {\bf 58} (2022), no.~11 221, [\href{http://arxiv.org/abs/2203.03209}{{\tt arXiv:2203.03209}}].

\bibitem{Thorsson:1993bu}
V.~Thorsson, M.~Prakash, and J.~M. Lattimer, {\it {Composition, structure and evolution of neutron stars with kaon condensates}},  Nucl. Phys. A {\bf 572} (1994) 693--731, [\href{http://arxiv.org/abs/nucl-th/9305006}{{\tt nucl-th/9305006}}]. [Erratum: Nucl.Phys.A 574, 851 (1994)].

\bibitem{Campbell:1974qt}
D.~K. Campbell, R.~F. Dashen, and J.~T. Manassah, {\it {Chiral Symmetry and Pion Condensation. 1. Model Dependent Results}},  Phys. Rev. D {\bf 12} (1975) 979.

\bibitem{Ellis:1995kz}
P.~J. Ellis, R.~Knorren, and M.~Prakash, {\it {Kaon condensation in neutron star matter with hyperons}},  Phys. Lett. B {\bf 349} (1995) 11--17, [\href{http://arxiv.org/abs/nucl-th/9502033}{{\tt nucl-th/9502033}}].

\bibitem{Tatsumi:1988up}
T.~Tatsumi, {\it {$K^-$on Condensation and Cooling of Neutron Stars}},  Prog. Theor. Phys. {\bf 80} (1988) 22.

\bibitem{Brown:1988ik}
G.~E. Brown, K.~Kubodera, D.~Page, and P.~Pizzochero, {\it {Strangeness Condensation and Cooling of Neutron Stars}},  Phys. Rev. D {\bf 37} (1988) 2042--2046.

\bibitem{Choi:2021ign}
K.~Choi, H.~J. Kim, H.~Seong, and C.~S. Shin, {\it {Axion emission from supernova with axion-pion-nucleon contact interaction}},  JHEP {\bf 02} (2022) 143, [\href{http://arxiv.org/abs/2110.01972}{{\tt arXiv:2110.01972}}].

\bibitem{Fischer:2021jfm}
T.~Fischer, P.~Carenza, B.~Fore, M.~Giannotti, A.~Mirizzi, and S.~Reddy, {\it {Observable signatures of enhanced axion emission from protoneutron stars}},  Phys. Rev. D {\bf 104} (2021), no.~10 103012, [\href{http://arxiv.org/abs/2108.13726}{{\tt arXiv:2108.13726}}].

\bibitem{Freedman:1977gz}
B.~Freedman and L.~D. McLerran, {\it {Quark Star Phenomenology}},  Phys. Rev. D {\bf 17} (1978) 1109.

\bibitem{Schertler:2000xq}
K.~Schertler, C.~Greiner, J.~Schaffner-Bielich, and M.~H. Thoma, {\it {Quark phases in neutron stars and a 'third family' of compact stars as a signature for phase transitions}},  Nucl. Phys. A {\bf 677} (2000) 463--490, [\href{http://arxiv.org/abs/astro-ph/0001467}{{\tt astro-ph/0001467}}].

\bibitem{Alford:2004pf}
M.~Alford, M.~Braby, M.~W. Paris, and S.~Reddy, {\it {Hybrid stars that masquerade as neutron stars}},  Astrophys. J. {\bf 629} (2005) 969--978, [\href{http://arxiv.org/abs/nucl-th/0411016}{{\tt nucl-th/0411016}}].

\bibitem{Arias-Aragon:future}
F.~Arias-Arag\'on and F.~Nola, {\it To be determined},  \href{http://arxiv.org/abs/27XX.XXXXX}{{\tt arXiv:27XX.XXXXX}}.

\bibitem{Takatsuka:2005bp}
T.~Takatsuka, S.~Nishizaki, Y.~Yamamoto, and R.~Tamagaki, {\it {Occurrence of hyperon superfluidity in neutron star cores}},  Prog. Theor. Phys. {\bf 115} (2006) 355--379, [\href{http://arxiv.org/abs/nucl-th/0601043}{{\tt nucl-th/0601043}}].

\bibitem{Vidana:2004rd}
I.~Vidana and L.~Tolos, {\it {Superfluidity of Sigma- hyperons in beta-stable neutron star matter}},  Phys. Rev. C {\bf 70} (2004) 028802, [\href{http://arxiv.org/abs/nucl-th/0405010}{{\tt nucl-th/0405010}}].

\bibitem{Potekhin:1997mn}
A.~Y. Potekhin, G.~Chabrier, and D.~G. Yakovlev, {\it {Internal temperatures and cooling of neutron stars with accreted envelopes}},  Astron. Astrophys. {\bf 323} (1997) 415, [\href{http://arxiv.org/abs/astro-ph/9706148}{{\tt astro-ph/9706148}}].

\bibitem{Mignani:2012mm}
R.~P. Mignani, D.~V. Putte, M.~Cropper, R.~Turolla, S.~Zane, L.~J. Pellizza, L.~A. Bignone, N.~Sartore, and A.~Treves, {\it {The birthplace and age of the isolated neutron star RX J1856.5-3754}},  Mon. Not. Roy. Astron. Soc. {\bf 429} (2013) 3517, [\href{http://arxiv.org/abs/1212.3141}{{\tt arXiv:1212.3141}}].

\bibitem{Ho:2006uk}
W.~C.~G. Ho, D.~L. Kaplan, P.~Chang, M.~van Adelsberg, and A.~Y. Potekhin, {\it {Magnetic Hydrogen Atmosphere Models and the Neutron Star RX J1856.5-3754}},  Mon. Not. Roy. Astron. Soc. {\bf 375} (2007) 821--830, [\href{http://arxiv.org/abs/astro-ph/0612145}{{\tt astro-ph/0612145}}].

\bibitem{Sartore:2012fk}
N.~Sartore, A.~Tiengo, S.~Mereghetti, A.~De~Luca, R.~Turolla, and F.~Haberl, {\it {Spectral monitoring of RX J1856.5-3754 with XMM-Newton. Analysis of EPIC-pn data}},  Astron. Astrophys. {\bf 541} (2012) A66, [\href{http://arxiv.org/abs/1202.2121}{{\tt arXiv:1202.2121}}].

\bibitem{Motch:2009nq}
C.~Motch, A.~M. Pires, F.~Haberl, A.~Schwope, and V.~E. Zavlin, {\it {Proper motions of thermally emitting isolated neutron stars measured with Chandra}},  Astron. Astrophys. {\bf 497} (2009) 423, [\href{http://arxiv.org/abs/0901.1006}{{\tt arXiv:0901.1006}}].

\bibitem{Hambaryan:2011bu}
V.~Hambaryan, V.~Suleimanov, A.~D. Schwope, R.~Neuhaeuser, K.~Werner, and A.~Y. Potekhin, {\it {Phase resolved spectroscopic study of the isolated neutron star RBS 1223 (1RXS J130848.6+212708)}},  AIP Conf. Proc. {\bf 1379} (2011), no.~1 195--196, [\href{http://arxiv.org/abs/1108.3897}{{\tt arXiv:1108.3897}}].

\bibitem{Tetzlaff:2011kh}
N.~Tetzlaff, T.~Eisenbeiss, R.~Neuhaeuser, and M.~M. Hohle, {\it {The origin of RXJ1856.5-3754 and RXJ0720.4-3125 -- updated using new parallax measurements}},  Mon. Not. Roy. Astron. Soc. {\bf 417} (2011) 617, [\href{http://arxiv.org/abs/1107.1673}{{\tt arXiv:1107.1673}}].

\bibitem{Hambaryan:2017wvm}
V.~Hambaryan, V.~Suleimanov, F.~Haberl, A.~D. Schwope, R.~Neuh{\"a}user, M.~Hohle, and K.~Werner, {\it {The compactness of the isolated neutron star RX J0720.4{\ensuremath{-}}3125}},  Astron. Astrophys. {\bf 601} (2017) A108, [\href{http://arxiv.org/abs/1702.07635}{{\tt arXiv:1702.07635}}].

\bibitem{Tetzlaff:2012rz}
N.~Tetzlaff, J.~G. Schmidt, M.~M. Hohle, and R.~Neuhaeuser, {\it {Neutron stars from young nearby associations the origin of RXJ1605.3+3249}},  Publ. Astron. Soc. Austral. {\bf 29} (2012) 98, [\href{http://arxiv.org/abs/1202.1388}{{\tt arXiv:1202.1388}}].

\bibitem{Pires:2019qsk}
A.~M. Pires, A.~D. Schwope, F.~Haberl, V.~E. Zavlin, C.~Motch, and S.~Zane, {\it {A deep XMM-Newton look on the thermally emitting isolated neutron star RX J1605.3+3249}},  Astron. Astrophys. {\bf 623} (2019) A73, [\href{http://arxiv.org/abs/1901.08533}{{\tt arXiv:1901.08533}}].

\bibitem{suzuki2021quantitative}
H.~Suzuki, A.~Bamba, and S.~Shibata, {\it Quantitative age estimation of supernova remnants and associated pulsars},  The Astrophysical Journal {\bf 914} (2021), no.~2 103.

\bibitem{zharikov2021psr}
S.~Zharikov, D.~Zyuzin, Y.~Shibanov, A.~Kirichenko, R.~Mennickent, S.~Geier, and A.~Cabrera-Lavers, {\it Psr b0656+ 14: the unified outlook from the infrared to x-rays},  Monthly Notices of the Royal Astronomical Society {\bf 502} (2021), no.~2 2005--2022.

\bibitem{IAXO:2019mpb}
{\bf IAXO} Collaboration, E.~Armengaud {\em et.~al.}, {\it {Physics potential of the International Axion Observatory (IAXO)}},  JCAP {\bf 06} (2019) 047, [\href{http://arxiv.org/abs/1904.09155}{{\tt arXiv:1904.09155}}].

\bibitem{Ayala:2014pea}
A.~Ayala, I.~Dom{\'\i}nguez, M.~Giannotti, A.~Mirizzi, and O.~Straniero, {\it {Revisiting the bound on axion-photon coupling from Globular Clusters}},  Phys. Rev. Lett. {\bf 113} (2014), no.~19 191302, [\href{http://arxiv.org/abs/1406.6053}{{\tt arXiv:1406.6053}}].

\end{thebibliography}\endgroup
\bibliographystyle{BiblioStyle}

\end{document}